\documentclass[sigconf,nonacm]{acmart}

\usepackage{user_custom}

\begin{document}

\title[\name]{\name: Private Neural Network Inference with Joint Optimization of Convolution and FHE Bootstrapping}

\author{{J}ae {H}yung {J}u}
\authornote{Both authors contributed equally to this research.}
\author{Jaiyoung Park}
\authornotemark[1]
\affiliation{%
  \institution{Seoul National University}
  \city{Seoul}
  \country{Republic of Korea}
}
\email{hpotato@snu.ac.kr}
\email{jypark@scale.snu.ac.kr}

\author{Jongmin Kim}
\affiliation{%
  \institution{Seoul National University}
  \city{Seoul}
  \country{Republic of Korea}}
\email{jongmin.kim@snu.ac.kr}

\author{Minsik Kang}
\affiliation{%
  \institution{Seoul National University}
  \city{Seoul}
  \country{Republic of Korea}}
\email{kaiser351@snu.ac.kr}

\author{Donghwan Kim}
\affiliation{%
 \institution{Seoul National University}
 \city{Seoul}
 \country{Republic of Korea}}
 \email{eastflame@snu.ac.kr}

\author{{J}ung {H}ee {C}heon}
\affiliation{%
  \institution{Seoul National University, \\CryptoLab Inc.}
  \city{Seoul}
  \country{Republic of Korea}}
\email{jhcheon@snu.ac.kr}

\author{{J}ung {H}o {A}hn}
\affiliation{%
  \institution{Seoul National University}
  \city{Seoul}
  \country{Republic of Korea}}
\email{gajh@snu.ac.kr}

\renewcommand{\shortauthors}{Jae Hyung Ju et al.}

\begin{abstract}
Fully homomorphic encryption (FHE) is a promising cryptographic primitive for realizing private neural network inference (PI) services by allowing a client to fully offload the inference task to a cloud server while keeping the client data oblivious to the server. 
This work proposes \name, an FHE-based solution for the PI of deep convolutional neural networks (CNNs). 
\name tackles the critical problem of the enormous computational cost for the FHE evaluation of CNNs.
We introduce a novel encoding method called Coefficients-in-Slot (CinS) encoding, which enables multiple convolutions in one HE multiplication without costly slot permutations.
We further observe that CinS encoding is obtained by conducting the first several steps of the Discrete Fourier Transform (DFT) on a ciphertext in conventional Slot encoding.
This property enables us to save the conversion between CinS and Slot encodings as bootstrapping a ciphertext starts with DFT.
Exploiting this, we devise optimized execution flows for various two-dimensional convolution (conv2d) operations and apply them to end-to-end CNN implementations.
\name accelerates the performance of conv2d-activation sequences by up to {5.68}$\times$ compared to state-of-the-art FHE-based PI work and performs the PI of a CNN at the scale of ImageNet within a mere few seconds.
\end{abstract}





\maketitle

\section{Introduction}
The advances in machine learning (ML) have opened up a vast pool of cloud-based services, even covering highly private areas such as healthcare~\cite{gulshan2016development, sharma2022dermatologist}, finance~\cite{addo2018credit}, and home surveillance systems~\cite{Bowditch_2020_surveillance}.
Such services can be provided simply by sending user data to a cloud server, which runs inference on proprietary neural networks.
However, such a method exposes the input user data to the service provider.
With the increasing concerns and the resultant regulations~\cite{eu_2016_gdpr, act_1996_health} on privacy, private inference (PI), which incorporates methods to perform neural network inference while 1) the server remains oblivious to the input user data and 2) the user obtains the inference result without getting any additional information of the proprietary neural network, is garnering immense attention.

Several cryptographic primitives enable PI, among which homomorphic encryption (HE) stands out due to its capability to perform a sequence of operations on encrypted data (\ie, ciphertexts) without any online user intervention.
When using HE, the user first generates several public keys required for computation in a one-time offline phase.
Once the server holds the generated public keys, the user only needs to perform the encryption of the input data and decryption of the final result.
The server is in charge of the whole computation process; it constructs an arithmetic circuit $\mathcal{C}$ for the inference task and evaluates the circuit using HE operations (HE ops) on the user's ciphertexts.

However, evaluating complex circuits such as convolutional neural network (CNN) inference, the main target of this work, is challenging due to the critical constraints of HE.
First, there exists a limit to the number of HE ops that can be performed on a ciphertext.
In a subclass of HE schemes referred to as fully HE (FHE), a special \emph{bootstrapping} operation resets the limit, enabling more HE ops on the ciphertext.
However, bootstrapping adds an enormous amount of computation to the circuit; thus, a large portion of FHE-based PI's execution time is spent on performing bootstrapping rather than useful HE ops.

Moreover, manipulating the order of encrypted data is costly. 
In HE, a vector is encrypted into a ciphertext and the only viable option for the server to reorder the data inside this vector is to perform an HE op that cyclically rotates the vector (HRot).
As HRot is not cheap, prior studies have developed numerous algorithms and data rearrangement methods to minimize the cost for data reorganization in convolutional layers~\cite{juvekar_2018_gazelle, lee_2022_access, lee_2022_low, huang_2022_cheetah, kim_2023_optcnn, Zhang_2021_gala}.

Despite these efforts, even the state-of-the-art FHE CNN implementations~\cite{lee_2022_access, lee_2022_low, kim_2023_optcnn} exhibit impractically long latency, taking 
minutes per inference for relatively simple problems such as CIFAR10/100 and MNIST. This hinders their adoption in real-world ML services.

Henceforth, we will refer to the evaluation of a convolutional layer as \textbf{conv2d} to distinguish it from a plain convolution ($*$) operation.

\subsection{Contribution}
\label{sec:intro:contrib}

In this paper, we propose \textbf{\name}, a set of algorithmic advancements on ring learning with errors (RLWE)-based FHE to evaluate neural network models. Specifically, we implement our algorithms on CKKS~\cite{cheon_2017_homomorphic}, a popular FHE scheme for machine learning.
\name incorporates a novel \emph{CinS encoding} method for CKKS ciphertexts, which enables an efficient conv2d algorithm (\S\ref{sec:intro:contrib:nested}). This encoding method, along with the conv2d algorithm, allows for the reformation of the bootstrapping circuit, which reduces the cost of bootstrapping and its adjacent operations by disassembling and reassembling the computational blocks in the circuit (\S\ref{sec:intro:contrib:joint}). Additionally, we reorder the components within a CNN layer in a FHE-friendly manner to optimize the execution flows for various conv2d types (\S\ref{sec:intro:contrib:schedule}).
Overall, \name overcomes the limitations of FHE-based PI of CNNs by minimizing the computational overhead and enabling end-to-end inference of a complex CNN within a few seconds.

\begin{table*}
    \centering
    \caption{Comparison of FHE conv2d using various encoding methods. $N$ is the ring degree in FHE, $f$ is the width/height of kernels, $w$ is the width/height of images, and $C = C_{in} = C_{out}$ is the number of input/output channels that fit in N. dwconv2d denotes depthwise conv2d.}
    \begin{tabular}{ccccccc}
    \toprule
        Conv2d type & Encoding method & \# of PMult& \# of HRot& \# of elements in a msg& Level consumption & \# of plaintexts\\
    \midrule
        \multirow{3}{*}{Conv2d} & Slot encoding~\cite{juvekar_2018_gazelle} & $2f^2C$ & $2f^2+2C-4$ & $N/2$ & 1 & $2f^2C$\\
        & Coefficient encoding~\cite{kim_2023_optcnn} & $2C-1$ & $C-1$ & $N$ & 1 & $C$\\
        & \textbf{Ours (CinS encoding)}& $\mathbf{C}$ & $\mathbf{2\sqrt{C} - 2}$ & $\mathbf{N}$ & \textbf{0} & $\mathbf{C}$\\
        \midrule
        \multirow{3}{*}{Dwconv2d} 
        & Slot encoding~\cite{juvekar_2018_gazelle}\textsuperscript{1} & $2f^2$ & $2f^2-2$ & $N/2$ & 1 & $2f^2C$\\
        & Coefficient encoding~\cite{kim_2023_optcnn}\textsuperscript{1} & $2C-1$ & $C-1$ & $N$ & 1 & $C$\\
        & \textbf{Ours (CinS encoding)} & $\mathbf{1}$ & $\mathbf{0}$ & $\mathbf{N}$ & \textbf{1} & $\mathbf{C}$\\
    \bottomrule
    \multicolumn{7}{l}{\textsuperscript{1} Previous work did not include dwconv2d, so we independently implemented dwconv2d based on their encoding methods.}
    \end{tabular}
    \label{tab:conv_comparison}
\end{table*}

\subsubsection{CinS encoding and efficient conv2d algorithm}
\label{sec:intro:contrib:nested}

In CKKS, a message vector $\mathbf{m}$ is first encoded into a plaintext \pmsg, which is then encrypted into a ciphertext \cmsg.
The original encoding method in CKKS is known as \emph{Slot encoding}, which allows for element-wise multiplication ($\odot$ in Eq.~\ref{eq:encoding}) and addition between message vectors in the encrypted state.
The core part of Slot encoding involves performing an inverse discrete Fourier transform (IDFT) on $\mathbf{m}$.
PI studies~\cite{kim_2023_optcnn,huang_2022_cheetah} have later identified that when using \emph{Coefficient encoding}, which skips the IDFT, encrypted multiplication (HMult) of ciphertexts results in convolution ($*$ in Eq.~\ref{eq:encoding}) between the messages.
Previously, with Slot encoding, HRot operations were necessary to aggregate element-wise multiplication results within a ciphertext. Coefficient encoding reduces the need for HRot operations in conv2d processes, thereby optimizing computational efficiency.

However,
even the Coefficient encoding approach faces significant inefficiencies due to two primary reasons.
First, it still requires heavy rotations after each conv2d operation.
Second, it cannot efficiently evaluate element-wise operations.
These limitations arise mainly due to HMult being equivalent to a global convolution across the entire vectors, disregarding the actual convolution length required for a specific conv2d.

To mitigate this issue, we introduce a new encoding, a {Coefficients-in-Slot (CinS)} encoding, best suited to implement various conv2d evaluations with the CKKS scheme. 
Our CinS encoding removes the inefficiency by making HMult result in a partial local convolution for each evenly-partitioned slice of the vectors; i.e., given $\mathbf{m} = (\mathbf{m}_0 | \mathbf{m}_1 | \cdots | \mathbf{m}_{C-1})$ and $\mathbf{m}^\prime = (\mathbf{m}_0^\prime | \mathbf{m}_1^\prime | \cdots | \mathbf{m}_{C-1}^\prime)$ for  
$\mathbf{m}_i, \mathbf{m}_i'$ in a subring, we obtain 
\begin{align*}
    &\mathpt{\mathbf{m}}_\text{CinS} \cdot \mathpt{\mathbf{m}^\prime}_\text{CinS} = \mathpt{\mathbf{m} *_\text{local} \mathbf{m}^\prime}_\text{CinS}\\
&=\langle\mathbf{m}_0 *_\text{part} \mathbf{m}_0^\prime | \mathbf{m}_1 *_\text{part} \mathbf{m}_1^\prime | \cdots | \mathbf{m}_{C-1}  *_\text{part} \mathbf{m}_{C-1}^\prime\rangle_\text{CinS},
\end{align*} {where $*_\text{part}$ denotes the convolution of two polynomials within a subring of the given plaintext space.

We note that the conventional encodings provides the following HE ops: }
\begin{equation}
\label{eq:encoding}
\begin{split}
\text{Slot encoding:}&\ \mathpt{\mathbf{m}}_\text{slot} \cdot \mathpt{\mathbf{m}^\prime}_\text{slot} = \mathpt{\mathbf{m} \odot \mathbf{m}^\prime}_\text{slot}\\
\text{Coefficient encoding:}&\ \mathpt{\mathbf{m}}_\text{coeff} \cdot \mathpt{\mathbf{m}^\prime}_\text{coeff} = \mathpt{\mathbf{m} * \mathbf{m}^\prime }_\text{coeff}.
\end{split}
\end{equation}

Based on the CinS encoding method, we can flexibly select the size of a slice in the vector according to the required convolution length of a specific conv2d and develop dedicated conv2d algorithms.
Our conv2d proposal using CinS encoding combines the best of both worlds: Coefficient encoding and Slot encoding. In our conv2d algorithm, multiplication induces rotation-less convolution in each slice between input images and kernels as in Coefficient encoding, while rotation induces partial sum of intermediate convolution result as in Slot encoding. 
Our optimized conv2d algorithms significantly reduce the computational complexity, performing fewer multiplications and rotations (see Table~\ref{tab:conv_comparison}). 

We note that \cite{cheon_2018_bootstrapping} also suggested leveraging a subring structure to deal with a small number $\ell$ of slots rather than $N/2$ full slots,
which is a widely adopted technique to reduce bootstrapping latency for sparsely-packed ciphertext in CKKS.
Compared to~\cite{cheon_2018_bootstrapping}, 
our CinS encoding leverages isomorphism in subring, which leads to favorable homomorphic property in conv2d while utilizing full $N/2$ slots.

\subsubsection{Fusing conv2d with boostrapping}
\label{sec:intro:contrib:joint}

Although CinS encoding is well-suited to conv2d, it must be converted to Slot encoding during CNN inference to perform element-wise operations, such as ReLU.
This conversion is achieved using the DFT ($\mathcal{T}$) and IDFT ($\mathcal{T}^{-1}$) matrices.
We first factorize the DFT matrix $\mathcal{T}$ into $\mathcal{T}_2^\prime \mathcal{T}_1^\prime$ using the Cooley-Tukey DFT factorization~\cite{cooley-tukey}.
Then, the conversion from Slot encoding to CinS encoding is possible by multiplying the ciphertext with $\mathcal{T}_1^\prime$; the opposite is possible by multiplying the ciphertext with $\mathcal{T}^{-1}\mathcal{T}_2^\prime$.

However, evaluating these matrix-vector multiplications is highly costly in the encrypted state.
Nevertheless, we identify that these evaluations are already present in the bootstrapping process, which enables us to merge conv2d with bootstrapping to perform the conversions at no additional cost.
We go even further by eliminating the need for $\mathcal{T}_2^\prime$ multiplication by fusing the matrix with conv2d kernel weights using the property that the $\mathcal{T}_2^\prime$ matrix and the kernel weights share a similar structure.
As a result, the fused conv2d evaluation has the same cost as a standard conv2d evaluation; thus, $\mathcal{T}_2^\prime$ multiplication becomes effectively free in the online phase.

\subsubsection{FHE-friendly execution flows}
\label{sec:intro:contrib:schedule}

In applying our conv2d algorithms to end-to-end CNN inference, we carefully design the execution flows such that the number of expensive operations (bootstrapping and HRot in particular) is minimized.
Flexibility is required for conv2d variants, including downsampling conv2d and depthwise conv2d (dwconv2d)~\cite{chollet2017xception}.
Handling downsampling conv2d is especially important because it produces sparsely packed ciphertexts, severely degrading the throughput of HE ops.
It is possible to merge multiple sparsely packed ciphertexts into a densely packed ciphertext, but the process involves a lot of bootstrapping.
To tackle this problem, we modify the execution flow by introducing a \emph{decomposed downsampling conv2d} algorithm, which involves $s^2$ times fewer bootstrapping operations for the stride $s$ of a downsampling conv2d.
We also devise an efficient dwconv2d algorithm that can be used along with CinS encoding to perform the dwconv2d evaluation with only a single HE op (see Table~\ref{tab:conv_comparison}).

\subsubsection{Implementation and evaluation}
\label{sec:intro:contrib:result}

We implemented the conv2d layers and representative CNN models, ResNet18/50, and MobileNetV2 for the ImageNet dataset, using \name. 
\name achieves up to 5.67$\times$ improved performance for the conv2d-activation sequence compared to prior state-of-the-art FHE-based CNN work~\cite{kim_2023_optcnn}.
For the ImageNet dataset, \name takes \textbf{5.35 seconds per inference} with a variant of \textbf{ResNet18} and \textbf{56.08 seconds} with \textbf{ResNet50}. This is remarkably fast considering that prior FHE-based CNN PI studies~\cite{kim_2023_optcnn, lee_2022_access, lee_2022_low} report minutes to hours of execution time even for smaller networks targeting CIFAR10/100.

\subsection{Related Work}
\label{subsec:related_work}
FHE-based PI offers robust privacy guarantees by ensuring that the client only learns the final inference result. Moreover, it effectively offloads the computational workload to the server, minimizing the client's role to encrypting and decrypting the input and output data, enhancing its usability in resource-constrained environments such as mobile devices. Additionally, this approach aligns well with hardware acceleration efforts, as evidenced by numerous studies exploring HE hardware acceleration studies~\cite{kim_2022_ark,kim_2022_bts, nikola_2022_craterlake, jung_2021_100x, 2020_riazi_heax}. While there exist alternative PI research directions with various cryptographic constructions \cite{liu_2017_minionn,juvekar_2018_gazelle,lehmkuhl_2020_delphi,rathee_2020_cryptflow2,ng_2021_gforce,huang_2022_cheetah, watson_2022_piranha}, we focus on FHE-based implementation to fully leverage the advantageous properties inherent to FHE.

The unique algebraic structure of HE prevents direct translation of conventional conv2d algorithms (e.g., im2col-based algorithms) into efficient HE algorithms.
To mitigate this issue, Gazelle~\cite{juvekar_2018_gazelle} proposes a HE-specific conv2d algorithm on Slot-encoded ciphertexts to minimize the number of HE ops (see Table~\ref{tab:conv_comparison}).
Following studies~\cite{Zhang_2021_gala, lee_2022_low, lee_2022_access} extend Gazelle's algorithm by improving on how multiple channels are packed in a ciphertext.
Meanwhile, Cheetah~\cite{huang_2022_cheetah} introduces a conv2d algorithm that does not require any HRot ops using the Coefficient-encoding method.
\cite{kim_2023_optcnn}, the state-of-the-art work in FHE-based PI which we set as our baseline, adapts Cheetah's conv2d algorithm to FHE circumstances and implements end-to-end inference of ResNet models.
HyPHEN~\cite{kim_2023_hyphen} goes beyond optimizing a single convolution to target an entire block composed of multiple layers (e.g., a residual block in ResNet). By expanding the design space, HyPHEN demonstrates a significant reduction in the number of HE operations and memory footprint, and introduces a novel trade-off between the number of bootstrapping operations and other HE operations. However, HyPHEN requires low-degree polynomial activation to be effective. 
Despite these advancements, significant inefficiencies still impede the practical adoption of FHE-based PI; \cite{kim_2023_optcnn} requires 368 seconds for the server to perform PI on a single CIFAR10 image with the ResNet20 model.
\section{Preliminaries}
\label{sec:background}


We represent vectors with bold lower-case letters (e.g., $\mathbf{m}$).
All vectors are column vectors.
$\lfloor \cdot \rfloor$, $\lceil \cdot \rceil$, and $\lfloor \cdot \rceil$ represent floor, ceiling, and rounding operations. $\langle \cdot \rangle$ and $[\cdot ]$ represent encoding and encryption.
\integer, \real, and \complex denote the set of integer, real, and complex numbers.
$\mathbb{Z}_q = \mathbb{Z}/q\mathbb{Z}$ is a ring of integers modulo $q$.
We denote a ($2N$)-th cyclotomic polynomial ring $\mathbb{Z}_q[X]/(X^N+1)$ by \ring, whose \emph{degree} $N$ is a power-of-two integer (typically, $2^{16}$). Table~\ref{tab:symbol_table} summarizes the notations we use in this paper.

\subsection{Homomorphic Encryption}
\label{sec:background:he}
\emph{Homomorphic encryption} (HE) is a set of encryption schemes that allow performing operations on encrypted data without decryption, and the result is correct with respect to the computation on plaintext. 
 An HE scheme is a collection of algorithms $HE=(KeyGen, Enc, Dec, Eval)$ with the following syntax:

 \begin{itemize}[leftmargin=*, noitemsep]
     \item KeyGen($1^\lambda$) $\rightarrow (\mathsf{pk},\mathsf{sk})$. Given the security parameter $\lambda$, the KeyGen algorithm outputs a public key \pk and a secret key \sk.
     \item Enc(\pmsg; \pk) $\rightarrow$ \cmsg. Given a public key \pk and a plaintext \pmsg as input, the encryption algorithm outputs a ciphertext \cmsg. 
     \item Dec(\cmsg; \sk) $\rightarrow$ \pmsg. Given a secret key \sk and a ciphertext \cmsg as input, the decryption algorithm outputs the plaintext \pmsg. 
     \item Eval($\mathcal{C}$; \ct{$\mathbf{m_1}$}, \dots, \ct{$\mathbf{m_k}$}) $\rightarrow$ \ct{$\mathbf{m'}$}. For an arithmetic circuit $\mathcal{C}$, the evaluation algorithm outputs a new ciphertext \ct{$\mathbf{m'}$} encrypting $\mathbf{m'}=\mathcal{C}(\mathbf{m_1}$, \dots, $\mathbf{m_k})$
 \end{itemize}


CKKS supports an arithmetic circuit $\mathcal{C}$ consisting of addition and multiplication on real and complex numbers.
The server performs HE ops to evaluate $\mathcal{C}$ on the user's input ciphertexts.
HE ops can be represented as OpName(\cmsg, $x$) \text{$\rightarrow$} \ct{$\mathbf{m'}$}, where $x$ can be a ciphertext (e.g., HAdd/HMult: homomorphic add/mult between ciphertexts), a plaintext (PAdd/PMult), or a constant (CAdd/CMult).
Finally, HRot homomorphically performs a cyclic rotation by an integer $r$ on the message (msg) elements, which can be represented as \ct{$(m_r, \dots, m_{\frac{N}{2}},m_1 \dots,m_{r-1})$}.
We denote HRot(\ct{$\mathbf{m}$}, $r$) to as \ct{$\mathbf{m}<\!<r$} and also \ct{$\mathbf{m}$}$<\!<r$ for convenience ($>\!>$ for the opposite direction).
Among these basic HE ops, HMult and HRot take 1--2 orders of magnitude longer computation time compared to the other basic HE ops on conventional platforms, thereby accounting for the majority of the execution time on most workloads.
For a formal description of the RNS-CKKS scheme, an optimized version of CKKS, please refer to the original paper~\cite{cheon_18_rns}.

\subsection{CKKS Encoding Methods}
\label{sec:background:packing}

Two methods exist for encoding a message vector $\mathbf{m} \in \mathbb{C}^{N/2}$ into a plaintext $\langle \mathbf{m} \rangle \in \mathcal{R}_q$: Slot encoding and Coefficient encoding.

\textbf{Slot Encoding.} 
We let $\zeta = \exp(\pi \sqrt{-1} /N)$ be a $(2N)$-th primitive root of unity and set $\zeta_j$ as $\zeta_j := \zeta^{5^{j}}$ for $0 \le j <N/2$. For a matrix
\[U = \begin{bmatrix}
        1 & \zeta_0 & \zeta_0^2 & \cdots & \zeta_0^{N-1} \\
        1 & \zeta_1 & \zeta_1^2 & \cdots & \zeta_1^{N-1} \\
        \vdots & \vdots & \vdots & \ddots & \vdots \\
        1 & \zeta_{N/2-1} & \zeta_{N/2-1} & \cdots & \zeta_{N/2-1}^{N-1} 
    \end{bmatrix}\in \mathbb{C}^{(N/2)\times N}, \]
Slot encoding outputs $\langle \mathbf{m} \rangle : = \sum_{i=0}^{N-1} f_i \cdot X^i$ such that $\mathbf{f} = (f_i)_{0 \le i < N} = \frac{1}{N} ( \overline{U}^{T} \cdot \mathbf{m} + U^{T} \cdot \overline{\mathbf{m}} )$. Then it satisfies that $\langle \mathbf{m} \rangle \cdot \langle \mathbf{m}^\prime \rangle = \langle \mathbf{m} \odot \mathbf{m}^\prime \rangle$ for element-wise vector multiplication $\odot$.
In ordere to convert convolution into element-wise multiplication, discrete Fourier transform (DFT) is utilized; decoding is equivalent to performing DFT on the coefficients of \pmsg to obtain $\mathbf{m}$ and encoding is equivalent to the reverse process, which is IDFT.
When using Slot encoding, element-wise addition (HAdd/PAdd), element-wise multiplication (HMult/PMult), and cyclic rotation of the msg $\mathbf{m}$ (HRot) is possible by HE ops.

\textbf{Coefficient Encoding.}  
We first let $\tilde{\mathbf{m}} = ({\sf Re}(\mathbf{m}) ~|~ {\sf Im}(\mathbf{m})) \in \mathbb{R}^{N}$, where ${\sf Re}(\mathbf{m})$ and ${\sf Im}(\mathbf{m})$ are vectors representing real and imaginary part of $\mathbf{m}$, respectively. We then define Coefficient encoding as setting $\tilde{\mathbf{m}} \in \mathbb{R}^{N}$ to be the coefficients of $\langle \mathbf{m} \rangle \in \mathcal{R}_q$.
Then, $\langle \mathbf{m} \rangle \cdot \langle \mathbf{m}^\prime \rangle = \langle {\tilde{\mathbf{m}} * \tilde{\mathbf{m}}^\prime} \rangle$ holds for (negacyclic) convolution $*$, defined in $\mathbb{R}^{N}$. 
Similar to Slot encoding, element-wise addition is possible; however, multiplication (HMult/PMult) results in convolution of the message elements.
Prior work exploits this convolution property to implement matrix-matrix and matrix-vector multiplications~\cite{kim_2023_optcnn,huang_2022_cheetah}.
HRot is not supported in Coefficient encoding.

We often use the notation $\mathpt{\cdot}_\text{slot}$ and $\mathpt{\cdot}_\text{coeff}$ to differentiate between the two encoding methods.
For both methods, the encoding results are multiplied by a large \emph{scale factor} $\Delta$ and are rounded to convert them into integer polynomials, which can then be embedded into $\mathcal{R}_q$.
However, as this follow-up step is not central to our discussion, we often omit it in this paper for brevity.

We make the following observations regarding the encoding methods, which we use extensively in the paper.
These properties can be extended to CinS encoding, which we explain in \S\ref{sec:nested}.

\begin{remarkk}
\label{remark:encoding}
We can interpret the same plaintext differently with regard to the encoding method; e.g., $\mathpt{\mathbf{m}}_\text{slot} = \mathpt{IDFT(\mathbf{m})}_\text{coeff}$ and $\mathpt{\mathbf{m}}_\text{coeff} = \mathpt{DFT(\mathbf{m})}_\text{slot}$.
\end{remarkk}
\begin{remarkk}
\label{remark:slot}
We can indeed interpret any HE op as being applied to slot-encoded ciphertexts.
For example, HMult for Coefficient encoding can be interpreted as $\langle \mathbf{m} \rangle_\text{coeff} \cdot \langle \mathbf{m}^\prime \rangle_\text{coeff} = \langle DFT(\mathbf{m}) \rangle_\text{slot} \cdot \langle DFT(\mathbf{m}^\prime) \rangle_\text{slot} = \langle DFT(\mathbf{m}) \odot DFT(\mathbf{m}^\prime) \rangle_\text{slot} = \langle DFT(\mathbf{m} * \mathbf{m}^\prime) \rangle_\text{slot} = \langle \mathbf{m} * \mathbf{m}^\prime \rangle_\text{coeff}$ based on the convolution theorem of DFT.
\end{remarkk}
\subsection{Decomposition of DFT matrix}
\label{sec:background:dft_decomposition}
We first define the special DFT matrix as 
\[V = \begin{bmatrix}
    1 & \zeta_0 & \cdots & \zeta_0^{N/2-1} \\
    1 & \zeta_1 & \cdots & \zeta_1^{N/2-1} \\
    \vdots & \vdots & \ddots & \vdots \\
    1 & \zeta_{N/2-1} & \cdots & \zeta_{N/2-1}^{N/2-1}
\end{bmatrix},\]
which is an $(N/2)\times (N/2)$ square matrix satisfying $U = [ V ~|~\sqrt{-1}~\cdot~V]$. Evaluating this structured square matrix $V$ and its inverse $V^{-1} =\frac{2}{N} \cdot \overline{V}^{T}$ plays a crucial role in the CKKS scheme, including Slot encoding. There is a line of work~\cite{chen2019improved, HDFT} to decompose the matrix $V$ into a product of sparse diagonal matrices for efficient matrix evaluation. The main idea is to permute the columns of $V$ using the $(N/2)\times (N/2)$ bit-reversal permutation matrix $\mathcal{P}$ with $\mathcal{P}^{-1}=\mathcal{P}$. After the permutation, $V$ can be decomposed. To be precise, we let $\omega_m = \exp( 2 \pi \sqrt{-1}/ m)$ be a $m$-th primitive root of unity. (In particular, $\omega_{2N} = \zeta$) and let $\mathcal{T}_n = \left( \omega_{4n}^{5^{i} \cdot {\sf rev}_n(j)}\right)_{0 \le i,j < n}$ for $n \le N/2$, where ${\sf rev}_n(j)$ denotes bit-reversal permutation of $j$ with length $n$. Then we have $\mathcal{T}_{N/2} = V \cdot \mathcal{P}$ and $\mathcal{T}_{n}$ is decomposed as follows:
\begin{align} \label{eq:decomp}
 \mathcal{T}_{n} = \begin{bmatrix}
    I_{n/2} & W_{n/2} \\
    I_{n/2} & -W_{n/2}
\end{bmatrix}\ \cdot \begin{bmatrix}
    \mathcal{T}_{n/2} & 0 \\
    0 & \mathcal{T}_{n/2} 
\end{bmatrix},
\end{align}
where $W_{n/2}$ denotes a diagonal matrix ${\sf diag}(\omega_{4n}^{5^i})_{0 \le i < n/2}$. Leveraging this property repeatedly, we can decompose $\mathcal{T}_{n}$ into $\log n$ number of matrices as:
\[\mathcal{T}_n = S_{n/2}^{(n)} \cdot S_{n/4}^{(n)} \cdots S_1^{(n)}, \ \ S_k^{(n)} = \begin{bmatrix}
    B_k & 0 & \cdots & 0 \\
    0 & B_k & \cdots & 0 \\
    \vdots & \vdots & \ddots & \vdots \\
    0 & 0 & \cdots & B_k
\end{bmatrix} \in \mathbb{C}^{n \times n}\]
where $S_k^{(n)}$ is composed of $(n/{2k})$ block diagonals of the matrix $B_{k} = \begin{bmatrix}
    I_{k} & W_k \\
    I_{k} & -W_k
\end{bmatrix} \in \mathbb{C}^{(2k)\times
(2k)}$. For $n=N/2$, we simply write $S_k^{(N/2)} = S_k$ for each $k$, such that
$\mathcal{T}_{N/2} = S_{N/4} \cdot S_{N/8} \cdots S_1$.
We refer to~\cite{HDFT} for a more detailed description.



\subsection{Bootstrapping of FHE}
\label{sec:background:boot}
Multiplication between $\Delta$-scaled polynomials produces $\Delta^2$-scaled polynomials.
\textbf{Rescaling} restores the scale factor to $\Delta$ by truncating $\Delta$ bits from the least significant bits (LSBs) from the coefficients of the polynomial.
In this process, the modulus of the ring decreases from $q$ to $q/\Delta$.
As the modulus cannot decrease indefinitely, we define the \emph{level} of a polynomial such that the modulus of the polynomial at level $i$ is defined as $\mathcal{Q}(i) = q / \Delta^{L - i}$ for the initial level $L$, $0 \leq i \leq L$, and $q = \mathcal{Q}(L) > \Delta^L$.
Rescaling reduces the level by one.


\textbf{Bootstrapping} is a process to recover the level of a ciphertext reduced by rescaling.
Here, we briefly explain the computational flow of bootstrapping and refer the readers to \cite{cheon_2018_bootstrapping, bossuat_2021_efficient, bossuat_2022_bootstrapping} for an explanation of the state-of-the-art CKKS bootstrapping algorithms.

The level of a ciphertext, $[\langle\mathbf{m}\rangle_\text{coeff}]$ for instance, reaches zero after $L$ rescaling.
If we change the modulus from $\mathcal{Q}(0)$ to $\mathcal{Q}(L)$ to continue operation, we get $[\langle\mathbf{m} + \sfrac{\mathcal{Q}(0)}{\Delta} \mathbf{l} \rangle_\text{coeff}]$, where $\mathbf{l}$ is an integer vector holding small values.
Bootstrapping removes the unwanted $\sfrac{\mathcal{Q}(0)}{\Delta} \mathbf{l}$ term by a complex sequence of HE ops.
First, we change the encoding method to Slot encoding by an operation called \emph{coefficient-to-slot} (CtoS) to utilize element-wise multiplication only available in Slot encoding.
CtoS is a homomorphic evaluation of multiplication with the matrix $V^{-1}$, where $V$ is the special DFT matrix defined previously.

Then, through \emph{modular reduction evaluation} (ModEval) operation, it is possible to obtain $[\langle\mathbf{m}\rangle_\text{slot}]$ from $[\langle\mathbf{m} + \sfrac{\mathcal{Q}(0)}{\Delta} \mathbf{l} \rangle_\text{slot}]$. 
Finally, we return to the original encoding method by performing \emph{slot-to-coefficient} (StoC), which is a homomorphic evaluation of multiplication with the DFT matrix $V$.
Due to Remark~\ref{remark:encoding}, the same bootstrapping process applies to $[\langle\mathbf{m}\rangle_\text{slot}]$ equally by interpreting it as $[\mathpt{IDFT(\mathbf{m})}_\text{coeff}]$.

Due to its complexity, bootstrapping is an extremely expensive process that takes most of the execution time in FHE CKKS workloads, with CtoS and StoC operations comprising a significant portion of its runtime. To alleviate this burden, we use the bit-reversal DFT matrix $\mathcal{T}_{N/2}$ and its inverse $\mathcal{T}_{N/2}^{-1}$ instead of $V$ and $V^{-1}$ in StoC and CtoS, respectively.
Then, we can interpret CtoS and StoC all in Slot encoding based on Remark~\ref{remark:slot} to obtain
\begin{align*}
    &\text{CtoS}: \langle \mathbf{m} \rangle_\text{coeff} = \langle \mathcal{T}_{N/2} (\mathcal{P}\mathbf{m}) \rangle_\text{slot} \xmapsto{\mathcal{T}_{N/2}^{-1}} \langle \mathcal{P} \mathbf{m} \rangle_\text{slot}, \\
    &\text{StoC}: \langle \mathcal{P}\mathbf{m} \rangle_\text{slot} \xmapsto{\mathcal{T}_{N/2}} \langle \mathcal{T}_{N/2}(\mathcal{P} \mathbf{m}) \rangle_\text{slot} = \langle \mathbf{m} \rangle_\text{coeff},
\end{align*}
which implies that $\mathcal{P} \mathbf{m}$, rearranged from $\mathbf{m}$ in bit-reversal order, is encoded in slots instead.
We refer to~\cite{HDFT} for more details about the advantages of using $\mathcal{T}_{N/2}$.
The order of the slots after CtoS does not play any role in the bootstrapping.
Therefore, once the message $\mathbf{m}$ is given, we may assume that its Coefficient encoding $\langle \mathbf{m} \rangle_\text{coeff}$ is encoded in the given order, but its Slot encoding $\langle \mathbf{m} \rangle_\text{slot}$ is encoded in bit-reversal order for brevity of explanation.

As the resulting bootstrapping circuit consists of many HE ops, it consumes multiple levels ($L_\text{boot}$) and produces an output ciphertext at level $L' = L - L_\text{boot}$, leaving only a small number of levels on which to perform multiplication operations.




\subsection{Prior HE Implementations of CNNs}

\subsubsection{Handling conv2d}
\label{sec:background:conv2d}
Conv2d processes an input feature map $I\in\mathbb{R}^{C_{in}\times \iwidth \times \iwidth}$ comprising $C_{in}$ channels each of dimension $\iwidth \times \iwidth$.
For a predetermined stride ($s$) and padding, a set of kernels represented by $K\in \mathbb{R}^{C_{out} \times C_{in} \times \kwidth \times \kwidth}$ is used to produce the output feature map $O\in\mathbb{R}^{C_{out}\times \iwidth' \times \iwidth'}$. 
We denote the $\mathbf{n}$-th channel of $I$ as $I_n$, the kernel corresponding to the $\mathbf{n}$-th input and $\mathbf{m}$-th output channel as $K_{m,n}$, and the $\mathbf{m}$-th channel of $O$ as $O_m$.
Then, conv2d is formalized as
\begin{equation*}
O_m = \sum_{n=1}^{C_{in}}O_{m,n} = \sum_{n=1}^{C_{in}}K_{m,n} *_\text{2D} I_{n}\text{.}
\end{equation*}

\begin{figure}
\centering
\includegraphics[width=.999\columnwidth]{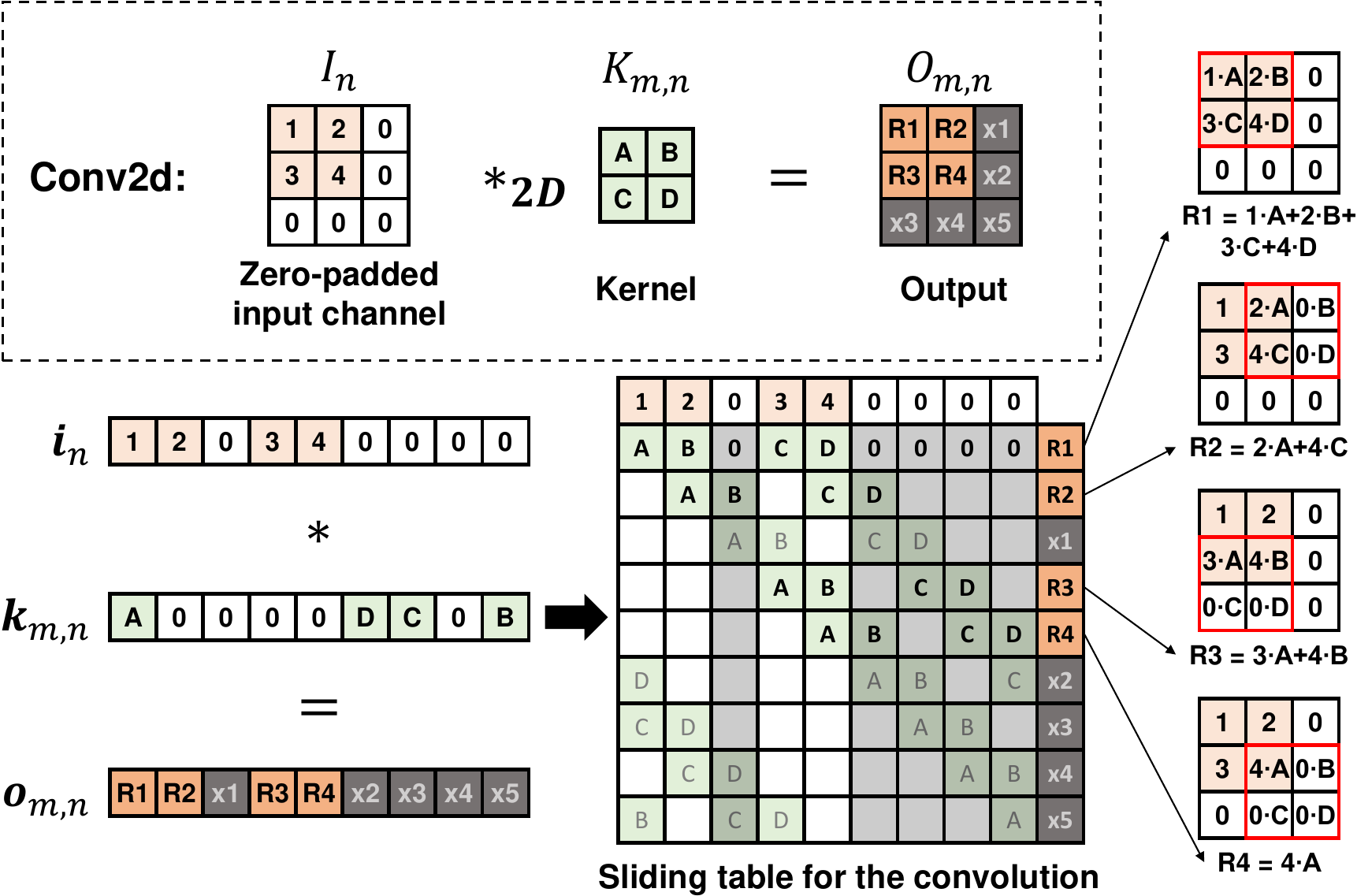}
\caption{Converting conv2d ($*_\text{2D}$) between a $2\times2$ input channel $I_n$ and a $2\times2$ kernel $K_{m, n}$, which produces a partial conv2d result $O_{m,n}$, into convolution between vectors: $\mathbf{i}_n * \mathbf{k}_{m,n} = \mathbf{o}_{m,n}$.}
\label{fig:conv2d-to-conv}
\end{figure}

Prior studies have identified that plain convolution can be utilized to perform higher-dimensional convolutions, such as $*_\text{2D}$, by zero-padding and flattening the data. This is shown in Figure~\ref{fig:conv2d-to-conv}, where $I_n$ and $K_{m,n}$ are converted into vectors $\mathbf{i}_n$ and $\mathbf{k}_{m,n}$. It can be observed from the figure that the convolution result $\mathbf{o}_{m,n} = \mathbf{i}_n * \mathbf{k}_{m,n}$ is equivalent to $O_{m,n}=K_{m,n} *_\text{2D} I_n$, flattened. Imitating higher-dimensional convolution (2D in the figure) using plain convolution creates unnecessary values (x1 to x5, painted black) in the output $O_{m,n}$. However these values appear precisely in the locations of zero-padding, and can be easily removed through an element-wise multiplication with a mask vector.

Based on this conversion from $*_\text{2D}$ to convolution, Cheetah~\cite{huang_2022_cheetah} proposes an efficient HE conv2d algorithm on Coefficient-encoded ciphertexts, on which our baseline~\cite{kim_2023_optcnn} improves.
For example, when $C = N / w^2$ (suppose $w$ is the padded input width/height and is a power-of-two number) channels can be placed in a ciphertext and $C = C_{in} = C_{out}$, the entire $I$ can be packed as a single ciphertext \ct{$\mathbf{i}$} and $\{K_{m,n} | n \in [1, C]\}$ can be packed together in a plaintext \pt{$\mathbf{k}_m$} for each $m$.
Prior studies devise a data organization method inside \ct{$\mathbf{i}$} and \pt{$\mathbf{k}_m$}'s for Coefficient encoding, such that PMult(\ct{$\mathbf{i}$}, \pt{$\mathbf{k}_m$}) results in a batch computation of $*_{2D}$ to finally produce a Coefficient-encoded ciphertext that encrypts the data of $O_m$ along with some unnecessary values.
The method is highly efficient compared to prior methods based on Slot encoding~\cite{juvekar_2018_gazelle}.



However, conv2d with Coefficient encoding still needs to gather $C$ ciphertexts each encrypting the data for $O_m$ ($m \in [1, C]$) into a single ciphertext that encrypts the entire $O$ for subsequent operations, incurring additional $C\text{-}1$ PMult and $C\text{-}1$ HRot ops for the data rearrangement.
Please refer to prior studies~\cite{juvekar_2018_gazelle, huang_2022_cheetah, kim_2023_optcnn} for the implementation details and cost analysis of each conv2d method; we summarize the final cost of the conv2d methods in Table~\ref{tab:conv_comparison}.





\subsubsection{Activation for FHE}
\label{sec:background:activation}
FHE only supports arithmetic circuits consisting of multiplication and addition, and non-polynomial functions for activation cannot be directly evaluated.
\cite{lee_2021_precise} proposes ReLU and MaxPool implementations based on high-degree polynomial approximation.
These approximation-based methods can directly replace the non-linear operations, but they incur huge computational overhead due to a lot of level consumption and frequent bootstrapping resulting from it.
For example, \cite{lee_2022_low} reports that over 83\% of the total inference time is spent on ReLU and bootstrapping.

To mitigate this overhead, \cite{park_2022_aespa, baruch_2023_sensitivetuning} propose retraining approaches that can construct CNNs with low-degree polynomial activation.
Contrary to the findings in \cite{garimella_2021_sisyphus, ishiyama_2020_highly, hesamifard_2017_cryptodl}, AESPA~\cite{park_2022_aespa} shows that, by adopting special training methods, deep CNNs can be effectively trained using low-degree polynomials.
With the retrained network, AESPA allows using a simple square function ($x^2$) for activation during inference
without compromising the inference accuracy.

The type and implementation of the activation function are orthogonal to our contributions.
We refer to the homomorphic evaluation of the activation function as HActivation$(\mathbf{ct})$ and only specify its type when presenting experiment results.
Our only assumption is that the input and output ciphertexts are slot-encoded for HActivation and that HActivation preserves the slot index of each value. 


\subsection{HE Matrix-Vector Multiplication Algorithm}\label{sec:background:diagonal pack}





We introduce a widely used matrix-vector multiplication method in HE~\cite{halevi2018helib}, which utilizes diagonal grouping.
Suppose the server wants to multiply a known $M\times M$ matrix $W$ to a length-$M$ vector encrypted as $[\langle \mathbf{m} \rangle_\text{slot}]$.
For the elements of $W$, $w_{i,j} (i,j \in [1,M])$, we refer to the vector of diagonally grouped elements starting with $w_{1,k}$ as $k$-th \emph{cyclic diagonal} of the matrix and write it as $\mathbf{w}_k^\text{diag}$.
Then, the multiplication of $W$ with $\mathbf{m}$ can be performed as the following:
\begin{equation}
\label{eq:matvecmul}
W\mathbf{m} = \sum_{k=1}^{M}\mathbf{w}_k^\text{diag}\odot(\mathbf{m}<\!<(k - 1))\text{.}
\end{equation}

When evaluating Eq.~\ref{eq:matvecmul} in the encrypted state, it requires $M - 1$ HRot and $M$ PMult ops.
As HRot ops are expensive, we can reduce the cost by using the baby-step giant-step (BSGS) algorithm ~\cite{halevi2018helib}.
For $B$ and $G$ that satisfies $B \cdot G = M$, BSGS modifies Eq.~\ref{eq:matvecmul} into a nested loop:
\begin{equation*}
\sum_{j=1}^G \left( \sum_{i=1}^B \mathbf{w}_{(j - 1)B + i}^{\text{diag}^\prime}\odot(\mathbf{m}<\!<(i - 1)) \right) <\!< (j - 1)B\text{.}
\end{equation*}
By reusing $m <\!< (i-1)$ values, the total number of HRot ops is reduced to $B + G - 2$; we can select $B = G = \sqrt{M}$ to minimize it to $2\sqrt{M} - 2$. The notation $\mathbf{w}_{(j - 1)B + i}^{\text{diag}^\prime}$ indicates $\mathbf{w}_{(j - 1)B + i}^{\text{diag}} >\!> (j - 1)B$. Since $W$ is known to the server, the plaintexts $\{\langle\mathbf{w}_{k}^{\text{diag}^\prime}\rangle_{\text{slot}}\}_{1 \le k \le BG}$ can be created in the preprocessing phase.

The BSGS algorithm can be extended for a sparse matrix with a few evenly-spaced cyclic diagonals.
For example, if a matrix has $M^\prime$ cyclic diagonals $\{\mathbf{w}_{1}^\text{diag}, \mathbf{w}_{1 + \ell}^\text{diag}, \cdots, \mathbf{w}_{1 + (M^\prime - 1)\ell}^\text{diag}\}$, BSGS can be applied to perform the matrix-vector multiplication with $2\sqrt{M^\prime} - 2$ HRot ops. 

The cost of matrix-vector multiplication can be further reduced if the matrix can be decomposed into a product of sparser matrices, as in \S\ref{sec:background:dft_decomposition}, where the BSGS algorithm can be applied for each decomposed sparse matrix.
As we decompose a matrix into a greater number of sparser matrices, the total number of HRot ops decreases at the cost of more level consumption.

The optimal point in this trade-off differs depending on the FHE parameters and the ciphertext level~\cite{chen2019improved}.
Deciding the optimal point is mostly orthogonal to our contribution, so we simply write HMatmul$(W, [\langle\mathbf{m}\rangle_\text{slot}])$ to represent a series of BSGS operations resulting in $[\langle W\mathbf{m}\rangle_\text{slot}]$.

\section{C\lowercase{in}S Encoding}\label{sec:nested}

We introduce CinS encoding, which is the central method for our efficient CNN implementation.
We recall that StoC in the bootstrapping process is a homomorphic evaluation of $\mathcal{T}_{N/2}$, the DFT evaluation with bit-reversal permutation, as described in \S\ref{sec:background:boot}. Then StoC converts the encoding type of $\langle \mathbf{m} \rangle_\text{slot} \in \mathcal{R}_q$ as 
\[ \text{StoC}: \langle \mathbf{m} \rangle_\text{slot} \mapsto \langle \mathcal{T}_{N/2}\cdot \mathbf{m} \rangle_\text{slot} = \langle \mathbf{m} \rangle_\text{coeff}.\]
We decompose $\mathcal{T}_{N/2}$ as $\mathcal{T}_{N/2} = \mathcal{T}'_2 \cdot \mathcal{T}'_1$ and propose an intermediate encoding between Slot encoding and Coefficient encoding as
\[ \langle \mathbf{m} \rangle_\text{CinS} := \langle \mathcal{T}_1' \cdot \mathbf{m} \rangle_\text{slot},\]
which would lead to $ \langle \mathcal{T}_2' \cdot (\langle \mathbf{m} \rangle_\text{CinS}) \rangle_\text{slot} = \langle \mathbf{m} \rangle_\text{coeff}$.
We then show that $\langle \mathbf{m} \rangle_\text{CinS}$ indeed contains the coefficient of each $\langle \mathbf{m}_i \rangle_\text{coeff}$ in its slots when given a proper concatenation $\mathbf{m} = (\mathbf{m}_{0}| \mathbf{m}_{1} | \cdots | \mathbf{m}_{C-1})$.

\begin{remarkk}\label{remark:SinC}
{One might consider the dual encoding method referred as Slots-in-Coefficient (SinC) encoding $\langle \mathbf{m} \rangle_\text{SinC}$, which contains the context about each $\langle \mathbf{m}_i \rangle_\text{slot}$ at its coefficients. 
However, we do not address the details of SinC encoding in this paper since we can not find any suitable applications for it.}
\end{remarkk}

\subsection{Exploiting Partial DFT}
\label{subsection:partialDFT}
We provide technical details for our CinS encoding method. Leveraging the decomposition of $\mathcal{T}_{N/2} $ described in \S\ref{sec:background:dft_decomposition}, we have
\[\langle \mathbf{m} \rangle_\text{coeff} =  \langle \mathcal{T}_{N/2}\cdot \mathbf{m} \rangle_\text{slot} = \langle S_{N/4} \cdot S_{N/8} \cdots S_1 \cdot \mathbf{m} \rangle_\text{slot}.\]
Suppose that we are given power-of-two integers $C, \ell$ with $N/2 = C \cdot \ell$. If we apply the decomposition property (Eq.~\ref{eq:decomp} in \S\ref{sec:background:dft_decomposition}) $\log \sfrac{N}{2\ell}$ times to the initial matrix $\mathcal{T}_{N/2}$, then we have
\begin{align*}
\mathcal{T}_{N/2} &= S_{N/4} \cdot S_{N/8} \cdots S_{\ell} \cdot \begin{bmatrix}
    \mathcal{T}_{\ell} & 0 & \cdots & 0 \\
    0 & \mathcal{T}_{\ell} & \cdots & 0 \\
    \vdots & \vdots & \ddots & \vdots \\
    0 & 0 & \cdots & \mathcal{T}_{\ell},
    \end{bmatrix}
\end{align*}
with $C$ number of block-diagonal matrices $\mathcal{T}_\ell$. Concatenating the message $\mathbf{m} \in \mathbf{C}^{N/2}$ as $\mathbf{m} = (\mathbf{m}_{0}~|~ \mathbf{m}_{1} ~|~ \cdots ~|~ \mathbf{m}_{C-1})$, then we have

\begin{align*}
    (S_{\ell/2} \cdots S_1) (\mathbf{m}) &= (S_{\ell}^{-1}\cdots S_{N/4}^{-1} \cdot \mathcal{T}_{N/2}) (\mathbf{m}) \\
    &= (\mathcal{T}_\ell(\mathbf{m}_{0}) ~|~ \mathcal{T}_\ell(\mathbf{m}_{1}) ~|~ \cdots ~|~ \mathcal{T}_\ell(\mathbf{m}_{C-1})).
\end{align*}

If we use a shorthand notation $S_{j \leftarrow i} = S_{j/2}S_{j/4}\cdots S_{i}$, one can write the bit-reversal DFT matrix as $\mathcal{T}_{N/2} = S_{N/2 \leftarrow \ell} \cdot S_{\ell \leftarrow 1}$. Now, we provide a formal definition of CinS encoding.

\begin{definition}[CinS Encoding]\label{def:nested} 
Let $C, \ell$ be power-of-two integers satisfying $N/2 = C \cdot \ell$. For a message vector $\mathbf{m}\in \mathbb{C}^{N/2}$ in bit-reversal order and a matrix decomposition $\mathcal{T}_{N/2} = S_{N/2 \leftarrow \ell} \cdot S_{\ell \leftarrow 1}$, we define a CinS encoding of $\mathbf{m}$ as 
\begin{align*}
\langle \mathbf{m} \rangle_{\text{CinS}_{\ell}} := \langle S_{\ell \leftarrow 1} \mathbf{m} \rangle_\text{slot}.
\end{align*}
\end{definition}

Here, the matrix $\mathcal{T}_{\ell}$ is obtained after permuting the columns in the DFT matrix of length $\ell$, denoted as $DFT_\ell$. We will often abbreviate $\langle \cdot \rangle_{\text{CinS}_\ell}$ as $\langle \cdot \rangle_\text{CinS}$ when the context $\ell$ is clear, and write $\mathcal{T}_\ell$ as $DFT_{\ell}$ for clarity. Then, we conclude the following theorem.

\begin{theorem}[Partial DFT Property]\label{thm:local} Let integers $C, \ell$, and a message $\mathbf{m}\in\mathbb{C}^{N/2}$ be given as described in Definition~\ref{def:nested}. If we concatenate the message $\mathbf{m}$ as $\mathbf{m} = (\mathbf{m}_{0}~|~ \mathbf{m}_{1} ~|~ \cdots ~|~ \mathbf{m}_{C-1})$, which is composed of $C$ slices having the length $\ell$, then we have
\begin{align*}
    \langle \mathbf{m} \rangle_\text{CinS} &= \langle (S_{N/2 \leftarrow \ell}^{-1} \cdot DFT) \cdot \mathbf{m} \rangle_\text{slot} \\
    &= \langle (\mathbf{\bhat{m}}_{0} ~|~ \mathbf{\bhat{m}}_{1} ~|~ \cdots ~|~ \mathbf{\bhat{m}}_{C-1}) \rangle_\text{slot},
\end{align*}
where $\mathbf{\bhat{m}}_{i} = DFT_{\ell} (\mathbf{m}_{i})$ for a partial DFT matrix $DFT_{\ell}$ of length $\ell$.
\end{theorem}



\subsection{Homomorphic Property of CinS Encoding}

We interpret CinS encoding from an algebraic perspective. From Theorem~\ref{thm:local}, CinS encoding of $\mathbf{m} = (\mathbf{m}_{0} | \mathbf{m}_{1} | \cdots | \mathbf{m}_{C-1}) \in (\mathbb{C}^{\ell})^{C}$ is
\begin{equation}\label{eq:nested_structure}
\begin{split}
    &\langle \mathbf{m} \rangle_\text{CinS} = \langle (\mathbf{\bhat{m}}_{0} ~|~ \mathbf{\bhat{m}}_{1} ~|~ \cdots ~|~ \mathbf{\bhat{m}}_{C-1}) \rangle_\text{slot} \\
    &= \langle (DFT_{\ell} (\mathbf{m}_{0}) ~|~ DFT_{\ell} (\mathbf{m}_{1}) ~|~ \cdots ~|~ DFT_{\ell} (\mathbf{m}_{C-1})) \rangle_\text{slot},
    \end{split}
\end{equation}
a concatenation of component-wise DFT vectors of $\mathbf{m}_i \in \mathbb{C}^{\ell}$. Here, we consider another cyclotomic polynomial ring $\mathbb{Z}_q[Y]/(Y^{2\ell}+1)$, denoted as $\mathcal{R}'_q$. Then we can regard CinS encoding $\mathbf{m}$ as $C$-tuples of plaintexts in $\mathcal{R}'_q$, each of which can be viewed as 
\[\langle DFT_\ell (\mathbf{m}_i) \rangle_\text{slot} = \langle \mathbf{m}_i \rangle_\text{coeff} \in \mathcal{R}'_q.\]
Letting $Y = X^{C}$, the ring $\mathcal{R}'_q = \mathbb{Z}_q[Y]/(Y^{2\ell}+1)$ can be regarded as a subring of $\mathcal{R}_q = \mathbb{Z}_q[X]/(X^{N}+1)$, and there exists a relation between two algebraic objects from the following theorem.

\begin{theorem}[Isomorphism] Let $\mathcal{R}_q$ and $\mathcal{R}_q'$ denote the $2N$-th and $2 \ell$-th cyclotomic polynomial ring, respectively, as described above. Then we have a module isomorphism as
\[\mathcal{R}_q = \mathbb{Z}_q[X]/(X^{N}+1) \simeq \left(\mathbb{Z}_q[Y]/(Y^{2\ell}+1)\right)^{C} =(\mathcal{R}'_q)^{C}. \]
\end{theorem}

\begin{proof}
        For an element $\mathbf{a} = \sum_{i=0}^{N-1} a_i X^i$ in $\mathcal{R}_q$, one can write it in the form of 
\[ \mathbf{a} = \sum_{i=0}^{N-1} a_i X^i = 
\sum_{k=0}^{C-1} \sum_{j=0}^{2\ell-1} a_{Cj+k} \cdot 
X^{Cj+k} 
= \sum_{k=0}^{C-1} \left( \sum_{j=0}^{2\ell-1} a_{Cj+k} \cdot Y^{j} \right) \cdot X^{k}.\]

If we set each $\mathbf{a}_{C,k}$ as $\mathbf{a}_{C,k} =\sum_{j=0}^{\ell-1} a_{Cj+k} \cdot Y^{j}\in\mathcal{R}'_q$, then the mapping $\mathbf{a} \mapsto (\mathbf{a}_{C,0}, \mathbf{a}_{C,1}, \cdots, \mathbf{a}_{C, C-1})$ induces an isomorphism $\mathcal{R}_q \simeq (\mathcal{R}'_q)^{C}$.
\end{proof}
From the view of Coefficient encoding, the plaintext $\langle \mathbf{m} \rangle_\text{coeff} \in \mathcal{R}_q$ is mapped onto $(\langle \mathbf{m}_0 \rangle_\text{coeff}, \cdots, \langle \mathbf{m}_{C-1} \rangle_\text{coeff}) \in (\mathcal{R}'_q)^{C}$. Thus, one may regard CinS encoding as a local Coefficient encoding, which can be viewed as 
\begin{align*}
    \langle \mathbf{m} \rangle_\text{CinS} &= \langle (\mathbf{\bhat{m}}_{0} ~|~ \mathbf{\bhat{m}}_{1} ~|~ \cdots ~|~ \mathbf{\bhat{m}}_{C-1}) \rangle_\text{slot} \\
    &= \langle (\mathbf{m}_{0} ~|~  \mathbf{m}_{1} ~|~ \cdots ~|~ \mathbf{m}_{C-1}) \rangle_\text{local-coeff}.
\end{align*}

\subsubsection{Local Convolution}
From the above observations, we now define a local convolution $*_\text{local}$ as follows.
For two message vectors $\mathbf{m} = (\mathbf{m}_{0} | \mathbf{m}_{1} | \cdots | \mathbf{m}_{C-1})$ and $\mathbf{m}^\prime = (\mathbf{m}_{0}^\prime | \mathbf{m}_{1}^\prime | \cdots | \mathbf{m}_{C-1}^\prime)$ given in bit-reversal order, we define a mapping $*_\text{local}$ as 
\begin{align*}
\mathbf{m} *_\text{local} \mathbf{m}' := 
(\mathbf{m}_{0} *_{\ell} \mathbf{m}'_{0} | \mathbf{m}_{1} *_{\ell} \mathbf{m}_{1}^\prime | \cdots | \mathbf{m}_{C-1} *_\ell \mathbf{m}_{C-1}^\prime),
\end{align*}
where $*_{\ell}$ denotes the negacyclic convolution defined in a subring $\mathcal{R}'_q = \mathbb{Z}_q[Y]/(Y^{\ell} +1)$.
It can be viewed as component-wise convolution $*_\ell$ between two elements lying in $(\mathcal{R}'_q)^{C}$. Thus, we have
\begin{align*}
&\langle \mathbf{m} *_\text{local} \mathbf{m}' \rangle_\text{CinS} \\
&= \langle (\mathbf{m}_{0} \!*_{\ell}\! \mathbf{m}'_{0} | \mathbf{m}_{1} \!*_{\ell}\! \mathbf{m}_{1}^\prime | \cdots | \mathbf{m}_{C-1} \!*_\ell\! \mathbf{m}_{C-1}^\prime) \rangle_\text{local-coeff}.
\end{align*}
Now, we show the homomorphic property of CinS encoding via the following theorem.

\begin{theorem}[Homomorphism for CinS encoding] For two message vectors $\mathbf{m}, \mathbf{m}'\in\mathbb{C}^{N/2}$, we have
\begin{align*}
\langle \mathbf{m} *_\text{local} \mathbf{m}^\prime \rangle_\text{CinS} = \langle \mathbf{m} \rangle_\text{CinS} \cdot \langle \mathbf{m}^\prime \rangle_\text{CinS}
\end{align*}
\label{homomorphism_theorem}
\end{theorem}
\begin{proof}
By the convolution theorem at subring $\mathcal{R}'_q$, we have ${\mathbf{m}_0 *_\ell \mathbf{m}_0^\prime} = \widehat{\mathbf{m}_0 *_\ell \mathbf{m}_0^\prime}$ for each $i$.
Utilizing the theorem, one can show the following:
\begin{equation}\label{eq:nested_proof}
\begin{split}
&\langle \mathbf{m} *_\text{local} \mathbf{m}^\prime \rangle_\text{CinS} \\
&= \langle ({\mathbf{m}_0 *_\ell \mathbf{m}_0^\prime} | {\mathbf{m}_1 *_\ell \mathbf{m}_1^\prime} | \cdots | {\mathbf{m}_{C-1} *_\ell \mathbf{m}_{C-1}^\prime}) \rangle_\text{local-coeff} \\
&= \langle (\widehat{\mathbf{m}_0 *_\ell \mathbf{m}_0^\prime} | \widehat{\mathbf{m}_1 *_\ell \mathbf{m}_1^\prime} | \cdots | \widehat{\mathbf{m}_{C-1} *_\ell \mathbf{m}_{C-1}^\prime}) \rangle_\text{slot}\\
&= \langle (\bhat{\mathbf{m}}_0 \odot \bhat{\mathbf{m}}_0^\prime | \bhat{\mathbf{m}}_1 \odot \bhat{\mathbf{m}}_1^\prime | \cdots | \bhat{\mathbf{m}}_{C-1} \odot \bhat{\mathbf{m}}_{C-1}^\prime) \rangle_\text{slot}\\
&= \langle (\bhat{\mathbf{m}_0} | \bhat{\mathbf{m}_1} | \cdots | \bhat{\mathbf{m}_{C-1}}) \rangle_\text{slot} \cdot \langle (\bhat{\mathbf{m}_0^\prime} | \bhat{\mathbf{m}_1^\prime} | \cdots | \bhat{\mathbf{m}_{C-1}^\prime}) \rangle_\text{slot}\\
&= \langle \mathbf{m} \rangle_\text{CinS} \cdot \langle \mathbf{m}^\prime \rangle_\text{CinS}
\end{split}
\end{equation}
    
\end{proof}

\subsubsection{Cyclic Rotation}
\label{subsubsec:cyclic_rotation_nested}
Another important functionality that CinS encoding supports is a cyclic rotation of the sequence of slices.
When interpreting a CinS-encoded ciphertext as a slot-encoded ciphertext following Eq.~\ref{eq:nested_structure}, we can perform HRot by a multiple of $\ell$ (e.g., $k\ell$) on the ciphertext and obtain 
$[\langle \mathbf{m} <\!< k\ell \rangle_\text{CinS}] = [\langle(\bhat{\mathbf{m}_{k + 1}} | \cdots | \bhat{\mathbf{m}_{C}} | \bhat{\mathbf{m}_{1}} | \cdots | \bhat{\mathbf{m}_{k}})\rangle_\text{slot}]$.

\subsection{CinS Encoding and Efficient Conv2d}
\label{sec:algorithm:conv}
Previous homomorphic conv2d algorithms, which utilizre HE ops provided by Slot and Coefficient encodings, are not well-suited to the computational patterns of conv2d. These methods incur additional data relocation steps involving an excessive amount of HRot ops.  
By leveraging the homomorphic properties of CinS encoding, we instantiate a more efficient conv2d algorithm, which results in much fewer HRot ops.

\subsubsection{Real CinS Encoding}
\label{sec:dense-pack}
We have explained the case where CinS encoding uses the message domain $\mathbb{C}^{N/2}$ in accordance with the message domain of Slot encoding. However, as most CNN models utilize only real values, half of the available space remains unused. To address this, we present a modification to the definition of CinS encoding, which effectively utilizes $\mathbb{R}^N$ as the message domain, enabling us to pack two times more data in a message.

For a length-$2\ell$ real vector $\mathbf{x}\in\mathbb{R}^{2\ell}$, we define a folded vector $\check{\mathbf{x}} \in \mathbb{C}^{\ell}$, whose $i$-th element $(\check{\mathbf{x}})_i$ is defined as $(\check{\mathbf{x}})_i=(\mathbf{x})_i+\sqrt{-1}(\mathbf{x})_{i+\ell}$. Then, for a real vector $\mathbf{m} = (\mathbf{m}_1 | \mathbf{m}_2 | \cdots | \mathbf{m}_C)\in\mathbb{R}^{N}$ composed of $C$ length-$2\ell$ slices, we redefine CinS encoding to be:

\begin{equation}
\label{eq:nested_real_vector}
\begin{split}
\langle \mathbf{m} \rangle_\text{CinS} &= \langle S_{\ell \leftarrow 1}\mathcal{P}_\ell(\check{\mathbf{m}}_1 | \cdots | \check{\mathbf{m}}_C) \rangle_\text{slot} \\
&= \langle(DFT'_{2\ell}(\mathbf{m}_1) | \cdots | DFT'_{2\ell}(\mathbf{m}_C))\rangle_\text{slot}
\end{split}
\end{equation}

Here, $\mathcal{P}_\ell$ is a block diagonal matrix of total size $N/2 \times N/2$, where each block is $P_\ell$, a bit-reversal permutation matrix of size $\ell \times \ell$. Then, the following property holds.

\begin{equation}
\label{eq:nested_real_vector2}
\begin{split}
&\langle ({\mathbf{m}_1} | \cdots | {\mathbf{m}_{C}}) \rangle_\text{CinS} \cdot \langle ({\mathbf{m}_1^\prime} | \cdots | {\mathbf{m}_{C}^\prime}) \rangle_\text{CinS} \\
=&\langle ({\mathbf{m}_1 *_{2\ell} \mathbf{m}_1^\prime} | \cdots | {\mathbf{m}_{C} *_{2\ell} \mathbf{m}_{C}^\prime}) \rangle_\text{CinS}\\
\end{split}
\end{equation}
Here, $DFT'_{2\ell}$ is a real-to-complex DFT~\cite{sorensen1987realfft} that accepts real vectors of length $2\ell$ and outputs complex vectors of length $\ell$. We present the proof in Appendix~\ref{app:real_cins_encoding_pf}. This explains why the number of elements in a message is $N$ for our methods in Table~\ref{tab:conv_comparison}.

\subsubsection{Data organization}
As we have described, CinS encoding can be utilized to perform convolution on multiple input slices of power-of-two length at once.
First, each ${\iwidth}_0\times {\iwidth}_0$-sized channel of the input feature map ($I_i$) is zero-padded to a $\iwidth \times \iwidth$ shape, where $\iwidth = 2^{\lceil \log_2 \iwidth_0 \rceil}$ is a power-of-two value.
The padded channel is flattened in the row-major order to form a vector $\mathbf{i}_i$.
Then, we can use CinS encoding to pack $C = N / \iwidth^2$ padded channels together in a ciphertext.
For the entire feature map, we need $\sfrac{C_{in}}{C}$ ciphertexts: 
\begin{equation*}
\{[\langle \mathbf{i}_{C(i - 1) + 1} | \mathbf{i}_{C(i - 1) + 2} | \cdots | \mathbf{i}_{Ci} \rangle_\text{CinS}]\ |\ i \in [1, \sfrac{C_{in}}{C}]\}\text{.}
\end{equation*}



For each of the $C_{out} \cdot C_{in}$ kernels of conv2d ($K_{i,j}$), we use the method in Figure~\ref{fig:conv2d-to-conv} to prepare a $\iwidth \times \iwidth$-sized vector $\mathbf{k}_{i,j}$, which can be directly used to perform convolution with the feature map.
Then, in the same way as the feature map, $C$ kernels can be packed into a ciphertext.
However, we first need to decide the order of kernel packing.
We make the following observation, which forms the core of our conv2d method:
\begin{remarkk}
\label{remark:matvecmul}
If we regard the feature map as a length-$C_{in}$ vector of channels $\mathcal{I}$ and conv2d weights as a $C_{out}\times C_{in}$ matrix of kernels $\mathcal{K}$, conv2d is equivalent to performing a matrix-vector multiplication $\mathcal{K}\mathcal{I}$, where the multiplication between a channel and a kernel is replaced by convolution between them.
\end{remarkk}

Based on Remark~\ref{remark:matvecmul} and inspired by the diagonal grouping method in \S\ref{sec:background:diagonal pack}, we devise a diagonal packing method for CinS-encoding-based conv2d.
We first present an example for a simple case when $C_{out} = C_{in} = C$.
$\mathbf{k}_{i,j}$ represents the padded kernel corresponding to the $i$-th ($i \in [1, C]$) input channel and the $j$-th ($j \in [1, C]$) output channel.
Then, we prepare $C$ plaintexts each encoding a cyclic diagonal from the matrix of $\mathbf{k}_{i,j}$'s; i.e., we prepare $\{ \langle\mathbf{k}^\text{diag}_k\rangle_\text{CinS} \ | \ k \in [1, C] \}$, where
\begin{equation*}
\mathbf{k}^\text{diag}_k = (\mathbf{k}_{1, k} | \mathbf{k}_{2, k + 1} | \cdots | \mathbf{k}_{C - k + 1, C} | \mathbf{k}_{C - k + 2, 1} | \cdots | \mathbf{k}_{C, k - 1})\text{.}
\end{equation*}

\subsubsection{Computation of conv2d}
\label{sec:algorithm:conv2d:basic}

With the prepared data organization, we can use a similar method to \S\ref{sec:background:diagonal pack} to perform conv2d when $C_{out} = C_{in} = C$.
For the input ciphertext $[\langle \mathbf{i} \rangle_\text{CinS}] = [\langle \mathbf{i}_1 | \mathbf{i}_2 | \cdots | \mathbf{i}_C \rangle_\text{CinS}]$, we compute
\begin{equation}
\label{eq:nested_conv}
\sum_{k = 1}^{C} \text{PMult}([\langle \mathbf{i} << (k-1)\iwidth^2 \rangle_\text{CinS}], \langle\mathbf{k}^\text{diag}_k\rangle_\text{CinS})\text{.}
\end{equation}
Eq.~\ref{eq:nested_conv} has the same computational flow as the matrix-vector multiplication introduced in \S\ref{sec:background:diagonal pack}.
This produces the conv2d output ciphertext $[\langle \mathbf{o} \rangle_\text{CinS}] = [\langle \mathbf{o}_1 | \mathbf{o}_2 | \cdots | \mathbf{o}_C \rangle_\text{CinS}]$ such that $\mathbf{o}_i = \sum_{j = 1}^C \mathbf{k}_{i, j} * \mathbf{i}_j$, which is what we want.
Eq.~\ref{eq:nested_conv} requires $C$ PMult and $C - 1$ HRot ops when computed na\"ively, but applying the BSGS algorithm reduces the number of HRot ops to $2\sqrt{C} - 2$ as shown in Table~\ref{tab:conv_comparison}.
As HRot dominates the execution time of conv2d, our conv2d algorithm roughly features $\sfrac{2}{(\sqrt{C} + 1)}$ times the complexity of the coefficient-packing-based algorithm, which incurs $C - 1$ HRot ops.

\subsubsection{Generalization to many channels}
\label{sec:algorithm:conv2d:general}

For a more general case where there are more input and output channels than $C$, we perform block matrix multiplication for the $\mathcal{K}\mathcal{I}$ matrix-vector multiplication. 
For example, when $C_{out} = C_{in} = 2C$,
\begin{equation*}
\mathcal{K}\mathcal{I} = \left(
    \begin{array}{c|c}
      \mathcal{K}_{11} & \mathcal{K}_{12}\\
      \hline
      \mathcal{K}_{21} & \mathcal{K}_{22}
    \end{array}
    \right) \left(
    \begin{array}{c}
      \mathcal{I}_1 \\
      \hline
      \mathcal{I}_2
    \end{array}
    \right) = \left(
    \begin{array}{c}
      \mathcal{K}_{11}\mathcal{I}_1 + \mathcal{K}_{12}\mathcal{I}_2 \\
      \hline
      \mathcal{K}_{21}\mathcal{I}_1 + \mathcal{K}_{22}\mathcal{I}_2
    \end{array}
    \right)\text{.}
\end{equation*}
By dividing $\mathcal{K}$ into size-$C\times C$ blocks and $\mathcal{I}$ into length-$C$ blocks, we can use the conv2d method for the $C_{out} = C_{in} = C$ case repeatedly to compute the final result.

\subsubsection{Depthwise Conv2d and Average Pooling}
\label{sec:algorithm:conv2d:depthwise}

Whereas dwconv2d with Coefficient encoding is performed in the same way as for regular conv2d, CinS encoding enables a more efficient implementation of dwconv2d due to its structure.
When $C_\text{out} = C_{in} = C$ as in \S\ref{sec:algorithm:conv2d:basic}, we can pack the entire kernels into a single plaintext as $\langle \mathbf{k}_1 | \mathbf{k}_2 | \cdots | \mathbf{k}_C \rangle_\text{CinS}$.
Then, dwconv2d is equivalent to simply performing a single PMult between the input ciphertext and the kernel plaintext.
Also, average pooling can be regarded as a special variant of downsampling dwconv2d where the kernels are equivalent across channels.

\section{Fusing Conv2d with Bootstrapping}
\label{sec:joint-evaluation}
As described in \S\ref{subsection:partialDFT}, conversions between Slot, CinS, and Coefficient encoding are inherent in FHE bootstrapping process.
Thus, incorporating efficient conv2d in CinS encoding leads to a favorable latency reduction. We show that the cost of StoC computation can further be reduced by jointly evaluating CinS conv2d with StoC.

\begin{figure}[t]
\centering
\includegraphics[width=.99\columnwidth]{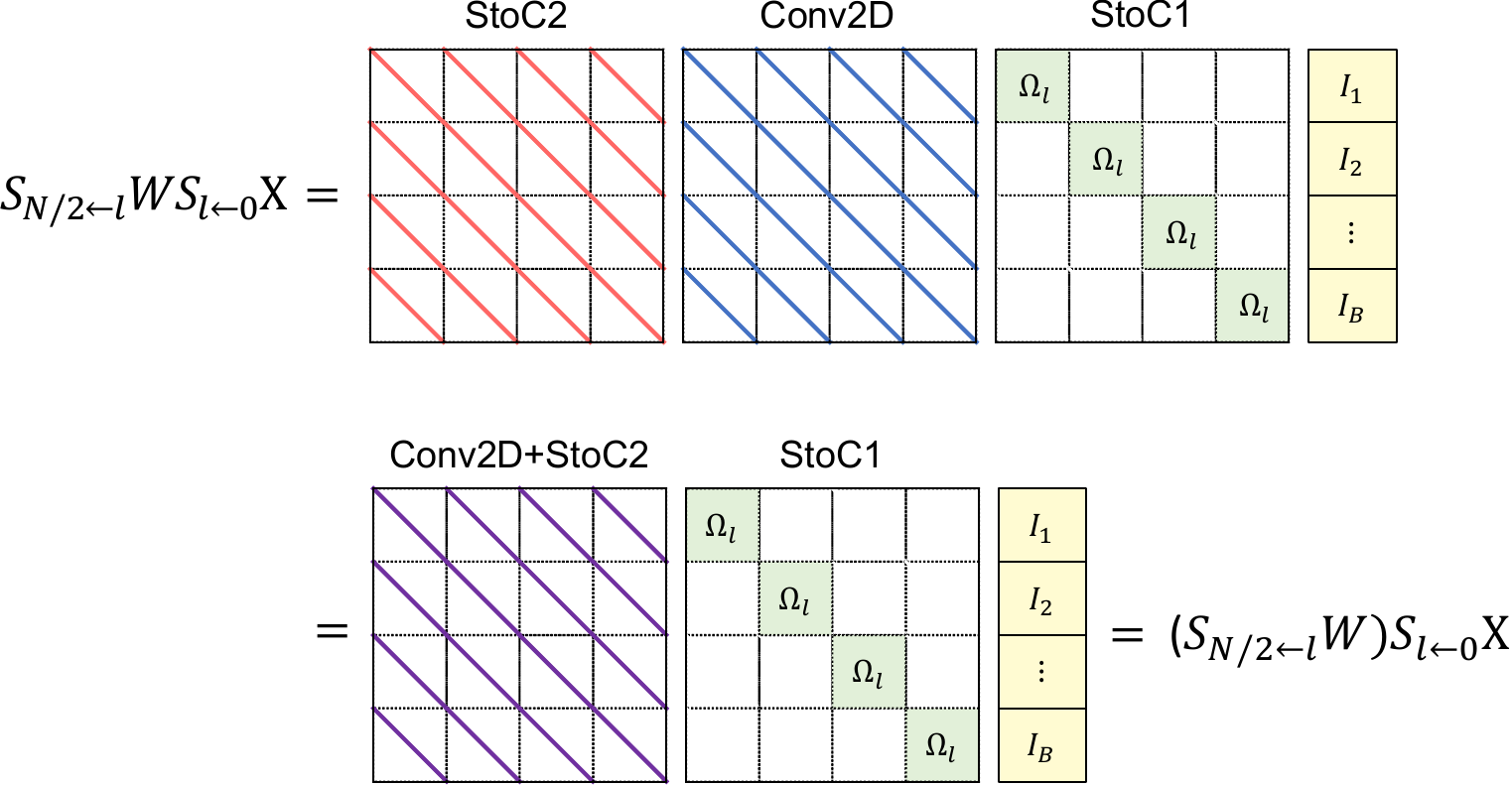}
\caption{Merging conv2D with StoC. The resulting composite operation simultaneously executes conv2D and StoC, and maintains the same computational cost as the original conv2D while canceling out the remaining StoC steps. Non-zero values exist only at the colored positions of the matrices. }
\label{fig:StoCPosteriorFusion}
\end{figure}

We optimize the operational flow by merging conv2d into the later part of StoC, StoC2 ($S_{N/2 \leftarrow \ell}$), to reduce the overall level consumption.
We first interpret conv2d operations on CinS-encoded ciphertexts as if being applied to slot-encoded ciphertexts based on Remark~\ref{remark:slot}.
We show that the StoC2 matrix $S_{N/2 \leftarrow \ell}$ can be fused with the conv2d kernel matrix ($\mathcal{K}$ in Remark~\ref{remark:matvecmul}) without increasing the amount of computation in terms of the number of HE ops for Slot-encoded ciphertexts.

\subsection{StoC2 matrix}
$S_{{N/2}\leftarrow{\ell}}$ corresponds to a set of butterfly operations with strides greater than or equal to $\ell$.
Therefore, the matrix $S_{{N/2}\leftarrow{\ell}}$ only operates in the granularity of a length-$\ell$ slice of the message vector.
In other words, for length-$\ell$ slices ${\mathbf{s}}_{i,j}$ that are composed of DFT twiddle factors and $\mathbf{m} = (\mathbf{m}_1 | \mathbf{m}_2 | \cdots | \mathbf{m}_C)$, the following holds:
\begin{equation}
(S_{{N/2}\leftarrow{\ell}}\mathbf{m})_i = \sum_{j=1}^{C}{{\mathbf{s}}_{i,j}\odot\mathbf{m}_j}\text{,}
\end{equation}
where $(S_{{N/2}\leftarrow{\ell}}\mathbf{m})_i$ is the $i$-th slice of $S_{{N/2}\leftarrow{\ell}}\mathbf{m}$ and $\mathbf{m}_j$ is the $j$-th slice of $\mathbf{m}$.
We can regard $S_{{N/2}\leftarrow{\ell}}$ as a $C\times C$ matrix $\mathcal{S'}$ with each element being a length-$\ell$ slice (${\mathbf{s}}_{i,j}$).
$\mathbf{m}$ can be similarly regarded as a length-$C$ vector $\mathcal{M}$ with each element being a length-$\ell$ slice ($\mathbf{m}_j$).
Then, $S_{{N/2}\leftarrow{\ell}}\mathbf{m}$ can be rewritten as a matrix-vector multiplication $\mathcal{S'}\mathcal{M}$ as in Remark~\ref{remark:matvecmul}, where the multiplication between slices are replaced by element-wise multiplication between them.

\begin{figure}[t]
\centering
\includegraphics[width=.8\columnwidth]{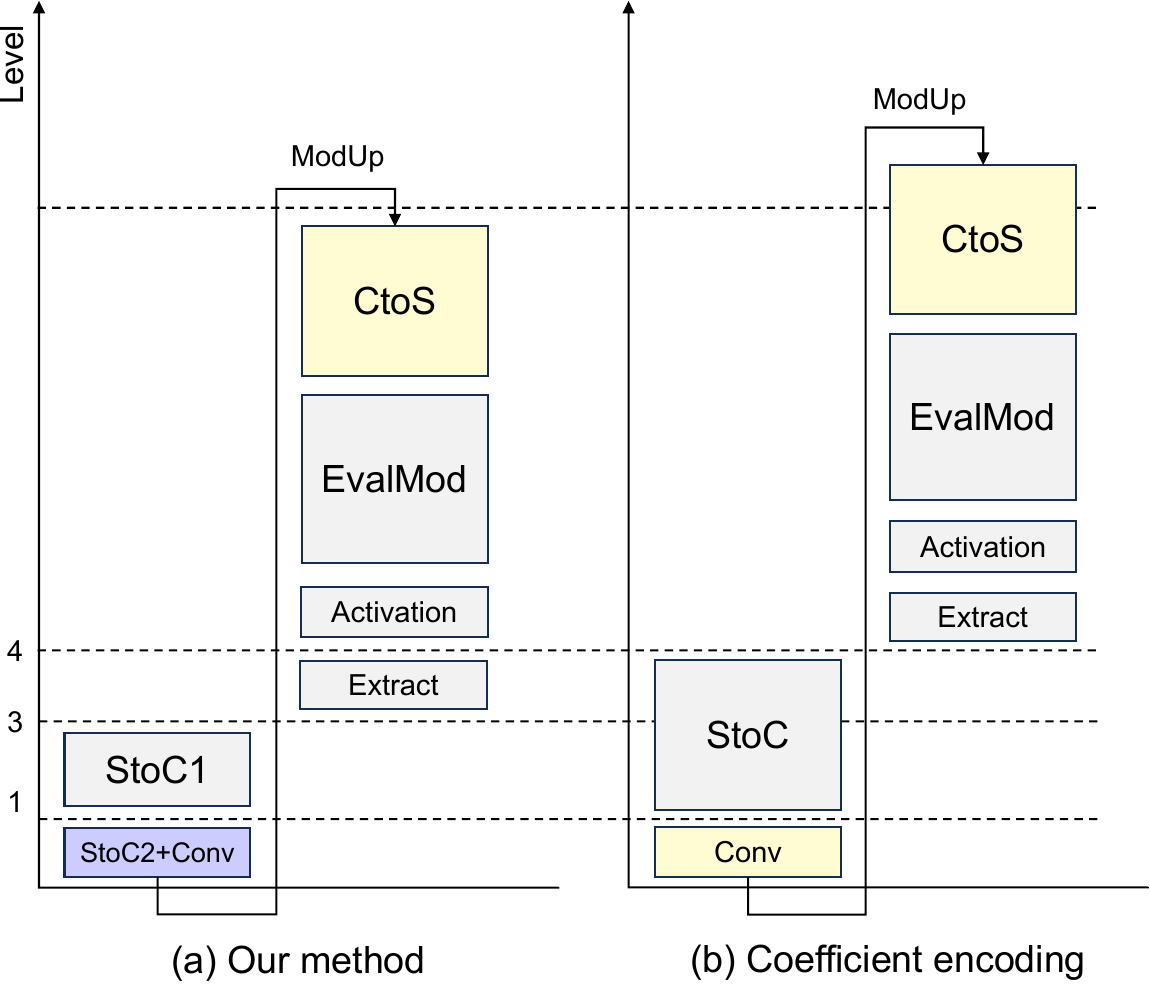}
\caption{Levels of ciphertexts when evaluating conv2d and activation along with boostrapping. We color operational blocks according to the encoding type of input ciphertexts: Slot-\colorbox{gray!20!white}{op}, Coefficient-\colorbox{yellow!20!white}{op}, and CinS-\colorbox{blue!20!white}{op}}
\label{fig:lowLevelDueToFusion}
\end{figure}

\subsection{Conv2d matrix}

If we reinterpret CinS-encoded ciphertexts as Slot-encoded, the slices $\mathbf{m}_j$ are converted to $\bhat{\mathbf{m}_j}$, and the local convolution $*_\text{local}$ is converted to element-wise multiplication $\odot$ (see Eq.~\ref{eq:nested_proof}).
As a result, Eq.~\ref{eq:nested_conv} is converted to the following equation in Slot encoding:
\begin{equation}
\bhat{\mathbf{o}_i} = \sum_{j=1}^{C}{\bhat{{\mathbf{k}}_{i,j}}\odot\bhat{\mathbf{i}_j}}\text{.}
\end{equation}

As in the previous section, we can regard this as a matrix-vector multiplication $\mathcal{\bhat{K}}\mathcal{\bhat{I}}$. $\mathcal{\bhat{K}}$ is a $C\times C$ matrix of length-$\ell$ slices ($\bhat{{\mathbf{k}}_{i,j}}$) and $\mathcal{\bhat{I}}$ is a length-$C$ vector of length-$\ell$ slices ($\bhat{\mathbf{i}_j}$).
Then, the process of computing conv2d and then StoC2 can be represented as the following:
\begin{equation}
\mathcal{S'}\mathcal{\bhat{K}}\mathcal{\bhat{I}} = (\mathcal{S'}\mathcal{\bhat{K}})\mathcal{\bhat{I}} = \mathcal{\bhat{K'}}\mathcal{\bhat{I}}\text{.}
\end{equation}

Linear algebra still holds even when each element in a matrix or a vector is replaced with a length-$\ell$ slice, and multiplication is replaced with element-wise multiplication between the slices.
Thus, the product of the two $C\times C$ matrices $\mathcal{S'}\mathcal{\bhat{K}}$ can be computed first.

As $\mathcal{S'}$ and $\mathcal{\bhat{K}}$ are composed of only twiddle factors and kernel weights, which are known to the server, the multiplication $\mathcal{\bhat{K}}^\prime =  \mathcal{S}^\prime\mathcal{\bhat{K}}$ can be precomputed by the server in the offline phase.
The resulting $\mathcal{\bhat{K'}}$ is a $C \times C$ matrix of length-$\ell$ slices, having the same form as $\mathcal{\bhat{K}}$.
Thus, the same computation process in \S\ref{sec:algorithm:conv2d:basic} can be used for the online phase.

That is, the cost of evaluating $\mathcal{S}^\prime\mathcal{\bhat{K}}\mathcal{\bhat{I}}$ is identical to the cost of evaluating $\mathcal{\bhat{K}}\mathcal{\bhat{I}}$, completely eliminating the cost of StoC2.
Hence, the cost of bootstrapping is significantly reduced by the fusion.

This fusion also reduces the level consumption by one and performs conv2d at the lowest level possible (level 1).
This minimizes the memory space required for storing kernel weights, as plaintext sizes are proportional to the level. Algorithm~\ref{alg:conv2d_default} describes the operational flow of our conv2d algorithm followed by activation.

We emphasize that fusion is only beneficial when the two matrices have the same granularity of $\ell$.
We provide an alternative explanation for the fusion in Figure~\ref{fig:StoCPosteriorFusion}.
In the figure, $\ell$-granular matrices are represented as a $C\times C$ matrix of blocks, where each block is shown as a $\ell \times \ell$ diagonal matrix, signifying element-wise operations on each slice.
Merging two $\ell$-granular matrices does not increase the number of cyclic diagonals, thereby retaining the evaluation cost equivalent to that of a single matrix.
It is not reasonable to additionally merge StoC1 because it will produce a dense matrix with a large number of non-zero cyclic diagonals, increasing the overall cost for StoC.

\subsubsection{Generalization of the fusion}

The fusion technique can be easily extended to encompass other linear layers within CNNs.
In particular, a fully-connected (FC) layer that performs matrix-vector multiplication utilizing a weight matrix, can be merged with the entire StoC matrix.
For FC layers, we use Slot encoding.
As Slot encoding is indeed equivalent to CinS encoding with a slice size of $\ell = 1$, the same formulation is applicable for the fusion process.
Also, for specific matrix shapes (e.g., $\mathcal{P}_\ell$ in Equation~\ref{eq:nested_real_vector}), we can even merge it with StoC1 and leave StoC2 for the use in conv2d.

\begin{algorithm}
\caption{Default Conv2d and Activation}\label{alg:conv2d_default}
\begin{algorithmic}[1]

\Require A ciphertext of $C$ $W\times W=2\ell$ sized input images with CinS encoding $\mathbf{ct}_\text{CinS, in}$, $C$ plaintexts of $C\times C$ preprocessed kernels with CinS encoding $\{ \langle\mathbf{k}^{\text{diag}^\prime}_i\rangle_\text{CinS}\}_{1 \le i \le C}$, two plaintexts to restore zero padding $\langle\mathbf{mask}_\text{Re}\rangle_\text{slot}$, $\langle\mathbf{mask}_\text{Im}\rangle_\text{slot}$

\Ensure A ciphertext of $C$ output images with CinS encoding $\mathbf{ct}_\text{CinS, out}$

\State $\mathbf{ct}_\text{coeff} \gets $BSGS$ (\mathbf{ct}_\text{CinS, in}, \{ \langle\mathbf{k}^{\text{diag}^\prime}_i\rangle_\text{CinS}\}_{1 \le i \le C}, \ell)$

\State $\mathbf{ct}_\text{slot,Re}, \mathbf{ct}_\text{slot,Im} \gets $ModEval$($CtoS$(\mathbf{ct}_\text{coeff}))$ 

\State $\mathbf{ct}_\text{slot,Re} \gets $HActivation$($PMult$(\mathbf{ct}_\text{slot,Re}, \langle\mathbf{mask}_\text{Re}\rangle_\text{slot}))$ 
\State $\mathbf{ct}_\text{slot,Im} \gets $HActivation$($PMult$(\mathbf{ct}_\text{slot,Im}, \langle\mathbf{mask}_\text{Im}\rangle_\text{slot}))$ 

\State $\mathbf{ct}_\text{slot} \gets $HAdd$(\mathbf{ct}_\text{slot,Re},$ iMult$(\mathbf{ct}_\text{slot,Im}))$ 

\State $\mathbf{ct}_\text{CinS, out} \gets $HMatmul$(S_{{\ell} \leftarrow 1}, \mathbf{ct}_\text{slot})$

\end{algorithmic}
\end{algorithm} 

\section{Execution Flow for Downsampling Conv2d}
\label{sec:flow:downsample}
\label{sec:flow}

Our proposed methods apply to any conv2d shape (e.g. pointwise conv2d), but special care is required for downsampling conv2d. We show how we rearrange downsampling conv2d to reduce the number of bootstrapping operations, and significantly reduce the overall computational cost.

\begin{algorithm}
\caption{Downsampling Conv2d and Activation}\label{alg:conv2d_downsample}
\begin{algorithmic}[1]

\Require Ciphertexts of $2C$ $W\times W=2\ell$ sized input images with Slot encoding $\mathbf{ct}_\text{slot, 1}$, $\mathbf{ct}_\text{slot, 2}$, $8C$ plaintexts of $4C\times 8C$ decomposed and preprocessed kernels with CinS encoding $\{ \langle\mathbf{k}^{\text{diag}^\prime}_i\rangle_\text{CinS}\}_{1 \le i \le 8C}$, two plaintexts to restore zero padding $\langle\mathbf{mask}_\text{Re}\rangle_\text{slot}$, $\langle\mathbf{mask}_\text{Im}\rangle_\text{slot}$

\Ensure A ciphertext of $4C$ $W/2\times W/2$ output images with CinS encoding $\mathbf{ct}_\text{CinS, out}$
\State $\mathbf{r} \gets [\ell/2, W/4, \ell/2+W/4]$
\For {$i \gets 1$ \textbf{to} $2$}
    \State $\mathbf{ct}_{\text{slot,ds},i} \gets $HMatmul$ (\mathcal{G}_{\ell}, \mathbf{ct}_{\text{slot},i})$   
        \For {$j \gets 1$ \textbf{to} $3$}
            \State $\mathbf{ct}_\text{slot,tmp} \gets $HMatmul$(\mathcal{G}_{\ell}, \mathbf{ct}_{\text{slot},i} <\!< \mathbf{r}[j])$
            \State $\mathbf{ct}_\text{slot,tmp} \gets \mathbf{ct}_\text{slot,tmp} >\!> (j\times \ell/4)$
            \State $\mathbf{ct}_{\text{slot,ds},i} \gets $HAdd$ (\mathbf{ct}_{\text{slot,ds},i}, \mathbf{ct}_\text{slot,tmp})$
        \EndFor
    \State $\mathbf{ct}_{\text{CinS},i} \gets $HMatmul$(S_{\ell/4 \leftarrow 1}, \mathbf{ct}_{\text{slot,ds},i})$
\EndFor

\State $\mathbf{ct}_{\text{coeff},1} \gets $BSGS$ (\mathbf{ct}_{\text{CinS},1}, \{ \langle\mathbf{k}^{\text{diag}^\prime}_i\rangle_\text{CinS}\}_{1 \le i \le 4C}, \ell/4)$
\State $\mathbf{ct}_{\text{coeff},2} \gets $BSGS$ (\mathbf{ct}_{\text{CinS},2}, \{ \langle\mathbf{k}^{\text{diag}^\prime}_i\rangle_\text{CinS}\}_{4C+1 \le i \le 8C}, \ell/4)$

\State $\mathbf{ct}_\text{coeff} \gets $HAdd$(\mathbf{ct}_{\text{coeff},1}, \mathbf{ct}_{\text{coeff},2})$

\State $\mathbf{ct}_\text{slot,Re}, \mathbf{ct}_\text{slot,Im} \gets $ModEval$($CtoS$(\mathbf{ct}_\text{coeff}))$ 

\State $\mathbf{ct}_\text{slot,Re} \gets $HActivation$($PMult$(\mathbf{ct}_\text{slot,Re}, \langle\mathbf{mask}_\text{Re}\rangle_\text{slot}))$ 
\State $\mathbf{ct}_\text{slot,Im} \gets $HActivation$($PMult$(\mathbf{ct}_\text{slot,Im}, \langle\mathbf{mask}_\text{Im}\rangle_\text{slot}))$ 

\State $\mathbf{ct}_\text{slot} \gets $HAdd$(\mathbf{ct}_\text{slot,Re},$ iMult$(\mathbf{ct}_\text{slot,Im}))$ 

\State $\mathbf{ct}_\text{CinS,out} \gets $HMatmul$(S_{\ell/4 \leftarrow 1}, \mathbf{ct}_\text{slot})$

\end{algorithmic}
\end{algorithm}

\begin{figure}[t]
\centering
\includegraphics[width=.97\columnwidth]{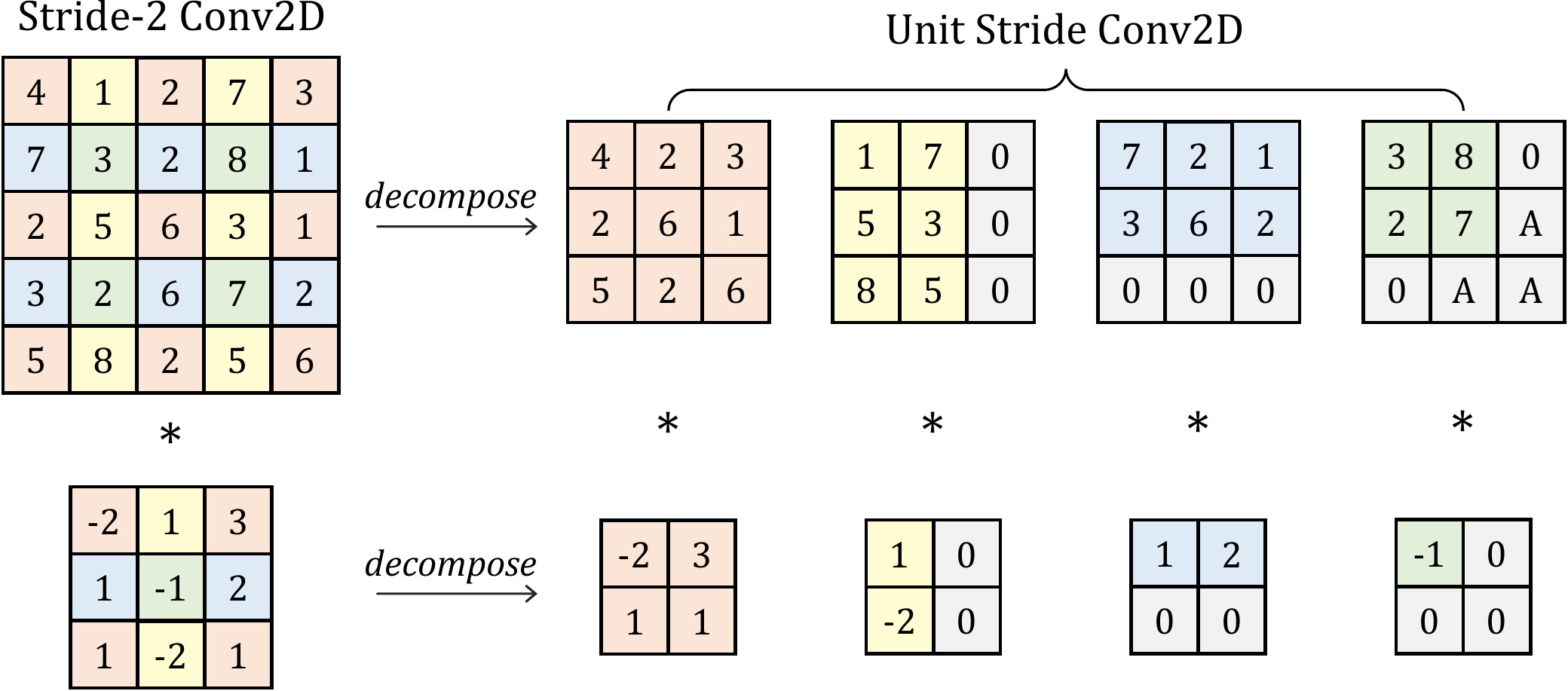}
\caption{Downsampling conv2d decomposition method for stride $s=2$.}
\label{fig:sconv-decomposition}
\end{figure}

Downsampling layers reduce the image size, and thus previous homomorphic conv2d algorithms output sparsely packed ciphertexts, which leads to slot underutilization. This can have a large impact on performance since the expansion of the number of ciphertexts necessitates a larger number of bootstrappings to process the same amount of data. Furthermore, a sparsely packed ciphertext cannot be reorganized to a densely packed ciphertext by simply changing memory addresses as in unencrypted data.
It requires a \emph{densify} operation, which is essentially an HMatmul op with a matrix ($\mathcal{G}_\ell$) that rearranges the scattered data into a sequential format. In more detail, the HMatmul consists of PMult ops with mask plaintexts that encode a vector composed of 0 and 1 to extract relevant slots, and many HRot ops to change the order~\cite{kim_2023_optcnn}. The resulting densify operation has high computational requirements and is often more expensive than the actual conv2d computation. We defer the details of $\mathcal{G}_\ell$ to Appendix~\ref{appendix:permutation_matrix}.

It becomes especially problematic when using the Coefficient or CinS encoding methods.
To perform the fine-grained HRot ops of densify, conversion to Slot encoding is required, resulting in the operational flow shown in Figure~\ref{fig:jobScheduling}(b).
This requirement forces bootstrapping on a large number of sparse ciphertexts.
For example, with a stride of $s=2$, a ciphertext stores only one valid value per four slots, thus four times more bootstrapping evaluations are required than necessary.

We devise a method to reduce the number of bootstrapping in downsampling conv2d by rearranging stride $s$ downsampling convolution into $s^2$ unit-stride conv2d operations as shown in Figure~\ref{fig:sconv-decomposition}.

\setlength{\tabcolsep}{5pt}
\begin{table*}[t]
    \centering
\caption{\centering 
\name inference time for ResNet18, ResNet50, and MobileNetV2 models using a single ImageNet image.
We implement the idea proposed in \cite{kim_2023_optcnn} and set it as our baseline. For ResNet18-AESPA, we also implemented the idea proposed in \cite{kim_2023_hyphen}. Polynomial approximation is used for activation functions, except in the case of ResNet18-AESPA. \label{tb:end-to-end}} 
    \begin{tabular}{c|ccc|cc|cc|cc}
        \toprule
        \textbf{Execution} & \multicolumn{3}{c|}{\textbf{ResNet18-AESPA}}& \multicolumn{2}{c|}{\textbf{ResNet18}} &\multicolumn{2}{c|}{\textbf{ResNet50}} & \multicolumn{2}{c}{\textbf{MobileNetV2}} \\
        \textbf{time (s)}  
            & \textbf{\cite{kim_2023_optcnn}} & \textbf{\cite{kim_2023_hyphen}} & \textbf{NeuJeans} & \textbf{\cite{kim_2023_optcnn}} & \textbf{NeuJeans} & 
            \textbf{\cite{kim_2023_optcnn}} & \textbf{NeuJeans} & \textbf{\cite{kim_2023_optcnn}} & \textbf{NeuJeans} \\ 
        \midrule
        \textbf{CtoS+ModEval}
            &10.57 & 3.27 &4.08 &14.23 &5.48  &33.71 &20.76&{\color{white} 0}23.24&16.70 \\
        \textbf{StoC}
            &0.68 & 0.35&0.46 &0.60 &0.48 &{\color{white} 0}2.63&{\color{white} 0}1.84&{\color{white} 0}1.85&1.68\\
        \textbf{Conv2d}
            &2.26 & 3.94&0.25 &2.30 &0.21 &{\color{white} 0}14.82&{\color{white} 0}1.92&{\color{white} 0}12.45&0.41\\
        \textbf{Activation} 
            &0.17 & 0.36&0.08 &7.55 &7.44 &{\color{white} 0}29.87&{\color{white} 0}30.61&{\color{white} 0}21.02&21.14\\
        \textbf{Densify/Decomp}
            &0.77 & 0.00&0.46 &1.22 &0.48 &{\color{white} 0}1.47 &{\color{white} 0}0.91&{\color{white} 0}1.01&0.79\\
        \textbf{ETC} 
            &0.02 & 1.82&0.02 &0.02 &0.02 &{\color{white} 0}0.03&{\color{white} 0}0.03&{\color{white} 0}0.04&0.01\\
        \midrule
        \textbf{Total}
            &14.47 & 9.74&5.35 &25.92 &14.11 & 82.53 & 56.08 &60.01&40.76\\
        \textbf{Speedup} 
            & - & $1.49\times$& $2.70\times$ & - & 1.84$\times$ & - & 1.47$\times$ & - & 1.47$\times$ \\ 
        \bottomrule
    \end{tabular}
\end{table*}
\setlength{\tabcolsep}{6pt}

As shown in Algorithm~\ref{alg:conv2d_downsample}, this method executes a decomposition step (lines 1-8), similar to densify, prior to conv2d. Afterwards, It converts each ciphertext to CinS encoding and performs unit-stride convolutions (line 9-12), and aggregates the convolution results to generate the densely packed ciphertexts embedding the downsampling convolution result (line 13).
Prepending the decomposition step significantly reduces the bootstrapping cost by facilitating an execution flow that avoids the need to bootstrap sparse ciphertexts.
The resulting execution flow is shown in Figure~\ref{fig:jobScheduling}(a). 
The decomposition initially resizes $w\times w$ images and $f\times f$ kernels to $s^2$ $\frac{w}{s} \times \frac{w}{s}$ images and $\frac{f}{s} \times \frac{f}{s}$ kernels.
During this process, the number of ciphertexts and plaintexts remains constant, as they are all densely packed, fully utilizing the slots. 

This modification has a large performance impact for complex networks like ResNet18 for ImageNet; we were able to reduce the number of total bootstrapping operations by more than two times for ResNet18.
We also further optimize densify/decompose operations by applying the BSGS optimization in \S\ref{sec:background:diagonal pack}.

\begin{figure*}[t]
    \centering
    \subfloat[Default conv2d]{\label{fig:ConvAESPA}\includegraphics[height=1.272in]{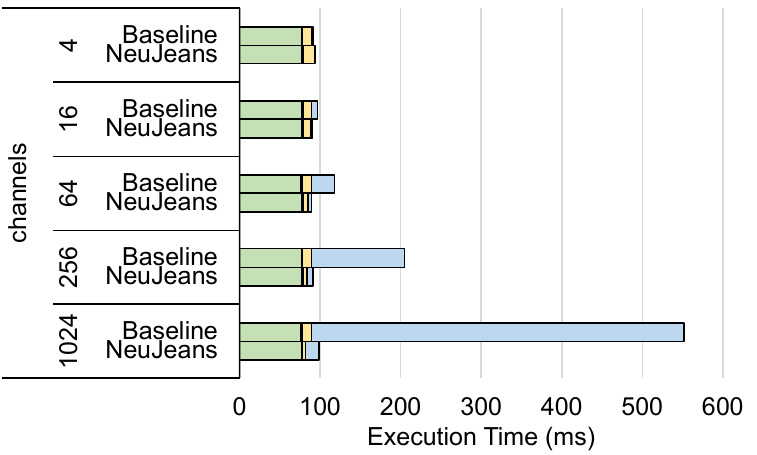}}
   \hspace{0.1in}
        \subfloat[Downsampling conv2d ($s=2$)]{\label{fig:Str_AESPA}\includegraphics[height=1.272in]{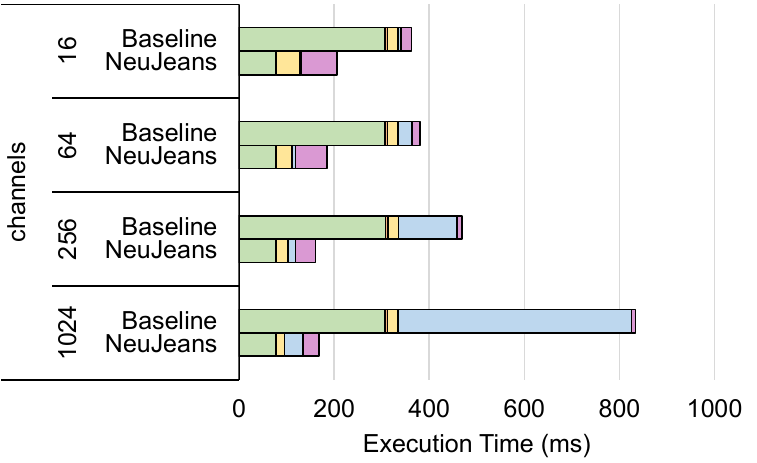}}
   \hspace{0.1in} 
    \subfloat[Depthwise conv2d]{\label{fig:DWAESPA}\includegraphics[height=1.272in]{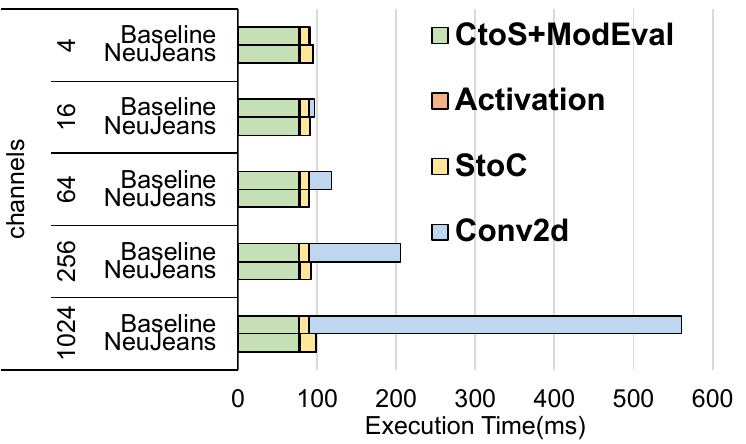}}
   \\
    \subfloat[Default conv2d]{\label{fig:ConvRELU}\includegraphics[height=1.272in]{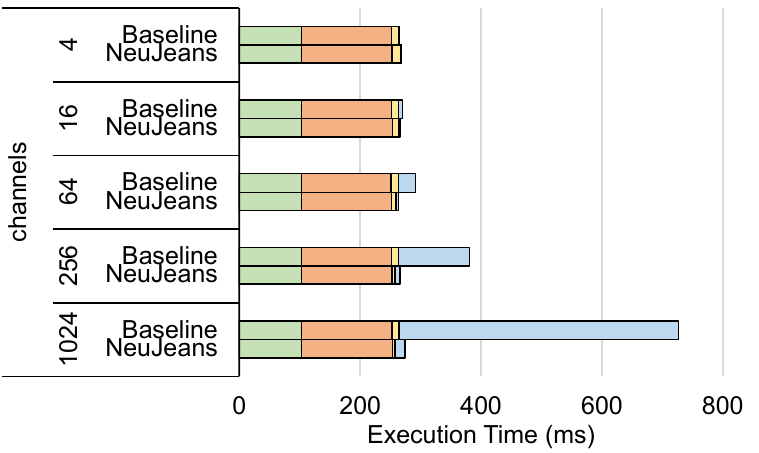}}
   \hspace{0.1in}
    \subfloat[Downsampling conv2d ($s=2$)]
    {\label{fig:StrRELU}\includegraphics[height=1.272in]{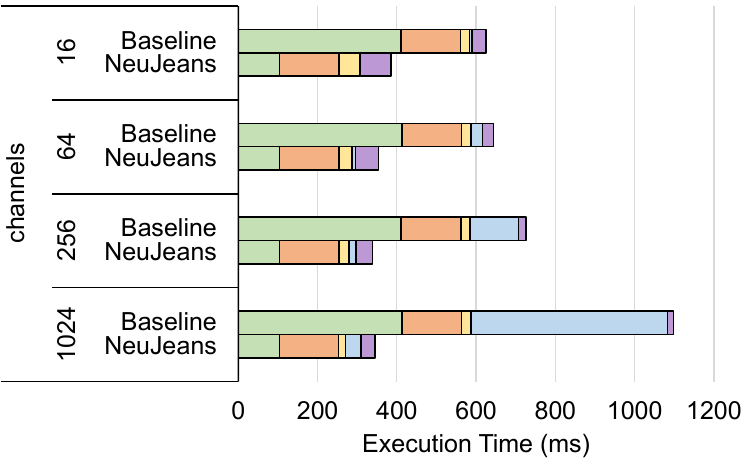}}
   \hspace{0.1in}
    \subfloat[Depthwise conv2d]{\label{fig:DWRELU}\includegraphics[height=1.272in]{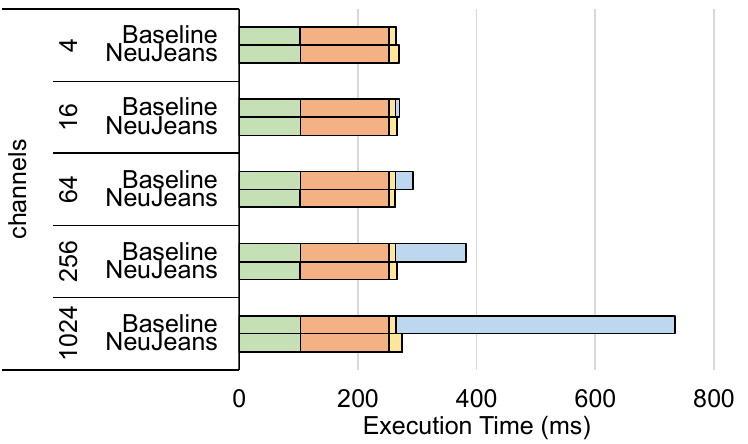}}
    
    \caption{Latency of default, downsampling ($s=2$), and depthwise conv2d followed by activation using the conv2d methods in baseline~\cite{kim_2023_optcnn} and \name for ResNet18 networks using the activation method of (a)(b)(c) AESPA and (d)(e)(f) polynomial approximation.}
    \label{fig:microbench}
\end{figure*}

\section{Evaluation}
\label{sec:evaluation}

\begin{table}
\centering
\caption{CKKS parameters used for the evaluation of \name. $N$ is the ring degree, $L$ is the max level, $\Delta$ is the scale factor for encoding, and $\lambda$ is the security level. 
}

\label{tab:param}
\begin{tabular}{c|ccccccc}
\toprule
Param set & $N$ & $L$ & $dnum$ & $\Delta$ & $h$ & $\log pq$ & $\lambda$\\
\midrule
Set1 & $2^{16}$ & 24 & 5 & $2^{42}$ & $192$ & $1555$ & $>128$\\
Set2 & $2^{16}$ & 19 & 4 & $2^{42}$ & $192$ & $1345$ & $>128$\\
\bottomrule
\end{tabular}
\end{table}

\subsection{Experimental Setup \& CNN Model Training}

We implemented \name using the HEaaN CKKS library with the support for GPU operation~\cite{heaan_library}. We select the parameters for CKKS as in Table~\ref{tab:param}, which guarantees over 128 bits of security~\cite{cheon_2021_fheparameter}.
We also implemented the state-of-the-art FHE-based PI methods proposed in ~\cite{kim_2023_optcnn, kim_2023_hyphen} using the same library to compare with \name. Specifically, we selected \cite{kim_2023_optcnn} as the representative method for coefficient encoding and \cite{kim_2023_hyphen} for slot encoding. However, \cite{kim_2023_hyphen} is not designed for CNNs that require high-degree polynomial approximations, thus we could not test it in such cases.
We measured the execution time of server-side tasks on a system with an AMD EPYC 7452 CPU, an NVIDIA A100 GPU, and 480GB of DDR4-3200 memory. Before execution, we load weight plaintexts and evaluation keys into GPU memory.
For the evaluation of client-side tasks, we used a laptop computer, Apple MacBook Air M1 with 8GB memory.
All client-side tasks use a single CPU thread and do not use GPU.

Using \name, we implemented ResNet18, ResNet50, and MobileNetV2 for the ImageNet dataset using ReLU \textbf{poly approx} (polynomial approximation) and additionally implemented ResNet18 using \textbf{AESPA}~\cite{park_2022_aespa}.
As described in \S\ref{sec:background:activation}, we can use a simple square function for activation after training with AESPA.
For AESPA, we trained the model for 200 epochs using standard supervised learning methods and employed the same hyperparameters as stated in the paper.
For poly approx, we used high-degree polynomial approximations of ReLU based on \cite{cheon_2020_effcomparision}.
We also attempted to employ ReLU approximation in \cite{lee_2022_minimax} but decided not to use it as it resulted in higher approximation errors and produced unstable results.
For all networks, MaxPool is replaced by AvgPool.
Whereas this replacement is handled within the training process for AESPA, we use the method described in \cite{baruch_2023_sensitivetuning} for poly approx, which gradually replaces MaxPool with AvgPool by fine-tuning a pre-trained model for 20 epochs.
We used the pre-trained network from \cite{wightman2021resnet} and trained the models using the PyTorch framework~\cite{paszke_2019_pytorch}.
%




\subsection{End-to-End CNN Inference Result}

\name reduces the execution time of FHE-based CNN inference to as low as 5.35 seconds, as shown in Table~\ref{tb:end-to-end}.
Compared to the baseline, \name achieves speedups of 1.84$\times$, 1.47$\times$, and 1.47$\times$ for ResNet18, ResNet50, and MobileNetV2 networks based on poly approx, while \name achieves 
2.70$\times$ speedup for ResNet18-AESPA network.
Larger speedups are achieved for AESPA because activation, which \name does not accelerate, constitutes a smaller portion in the execution time for AESPA.

The overall performance gains are attributed to 7.72--30.37$\times$ speedups on conv2d and 1.39--2.59$\times$ speedups on CtoS+ModEval compared to the baseline.
\name's conv2d algorithm based on CinS encoding leads to faster conv2d execution by significantly reducing the number of HE ops.
Also, our optimized execution flow for downsampling conv2d reduces the number of bootstrapping evaluations, which in turn decreases the CtoS+ModEval time.

Similar to \cite{kim_2023_optcnn}, \name performs conv2d and activation on different encodings, requiring bootstrapping for each conv2d-activation sequence to avoid additional costs from encoding transitions. In contrast, \cite{kim_2023_hyphen} exclusively uses slot encoding and performs bootstrapping after every two conv2d layers, thereby reducing the overall bootstrapping time. However, the significant number of rotations required for conv2d in \cite{kim_2023_hyphen} outweigh the benefits of the reduced bootstrapping time.
Consequently, \name achieves a 1.82$\times$ speedup for ResNet18-AESPA compared to \cite{kim_2023_hyphen}. 

\subsection{Microbenchmark Result}

\setlength{\tabcolsep}{6pt}
\begin{table}[t]
    \centering
\caption{\centering ImageNet image classification Top-1 and Top-5 Accuracy (\%) with 3000 tested images. \label{tb:Accuracy Analysis}} 
    \begin{tabular}{c|cc|cc}
        \toprule
        \multirow{3}{*}{\begin{tabular}{c}\textbf{Network}\\\textbf{type}\end{tabular}}&  \multicolumn{2}{c|}{\textbf{Unencrypted}} & \multicolumn{2}{c}{\textbf{NeuJeans}} \\
        & \multicolumn{2}{c|}{\textbf{accuracy}} & \multicolumn{2}{c}{\textbf{accuracy}} \\
        & \textbf{Top-1} & \textbf{Top-5} & \textbf{Top-1} & \textbf{Top-5} \\ 
        \midrule
        ResNet18-AESPA &66.4 &87.3&66.4 & 87.2 \\
        ResNet18 &67.2 &87.7&64.8&85.4 \\
        ResNet50 &76.4 &92.9&74.1&90.5 \\
        \bottomrule
    \end{tabular}
\end{table}
\setlength{\tabcolsep}{6pt}

To further analyze how \name accelerates conv2d, we conducted microbenchmarks for various types of conv2d.
Figure~\ref{fig:microbench} presents the results for default, downsampling, and depthwise conv2d, each followed by an activation function.
We used the baseline case of $C=C_{in}=C_{out}$, performing conv2d on a single ciphertext while varying $C$, the number of channels that can be packed into a ciphertext.
For downsampling conv2d, we used the case of $C=2C_{in}=C_{out}$, where two ciphertexts are downsampled to fit into a single ciphertext.
These results can be generalized to other cases based on \S\ref{sec:algorithm:conv2d:general}.

For default conv2d (Figures~\ref{fig:microbench}(a) and \ref{fig:microbench}(d)), we observe that the total execution time remains relatively constant when using \name, whereas the execution time of the baseline rapidly increases with $C$.
This is due to two main factors.
First, the HRot complexity of conv2d increases slowly, in proportion to $\sqrt{C}$ for \name, while it increases in proportion to $C$ for the baseline (see Table~\ref{tab:conv_comparison}).
Second, when many channels fit in a ciphertext, StoC time decreases by a large margin.
For the DFT matrix $\mathcal{T}_{N/2} = S_{N/2 \leftarrow \ell}S_{\ell \leftarrow 0}$, \name fuses $S_{N/2 \leftarrow \ell}$ with the evaluation of conv2d, leaving only $S_{\ell \leftarrow 1}$ multiplication for StoC.
For a large $C$, $\ell=N/C$ becomes small, which makes the evaluation of $S_{\ell \leftarrow 1}$ multiplication much cheaper.
Overall, up to 5.62$\times$ (respectively, 2.65$\times$) speedups are achieved for conv2d that uses the AESPA (poly approx) activation method.


In downsampling conv2d (Figure~\ref{fig:microbench}(b) and \ref{fig:microbench}(e)), the difference in execution time is more pronounced even for low-$C$ cases.
This is primarily due to the optimized decomposition-based execution flow for downsampling conv2d, which requires performing bootstrapping on only one-fourth the number of ciphertexts compared to the baseline.
As a result, CtoS+ModEval time decreases by 3.98--4.01$\times$.
Also, the aforementioned benefits of using \name for default conv2d apply in a similar manner for downsampling conv2d.


Finally, depthwise conv2d (Figure~\ref{fig:microbench}(c) and \ref{fig:microbench}(f)) shows a similar trend to default conv2d. 
In fact, the baseline performs an identical job for both cases.
Meanwhile, \name further reduces the execution time of depthwise conv2d by requiring only a single PMult op for its evaluation.
Overall, up to 5.68$\times$ (respectively, 2.68$\times$) speedups are achieved for conv2d that uses the AESPA (poly approx) method.


\subsection{Accuracy}

In Table~\ref{tb:Accuracy Analysis}, we present the FHE classification accuracy results for the ResNet18 and ResNet50 networks, based on 3,000 random samples from the ImageNet validation set, along with unencrypted classification results.
For the ResNet18 model trained using AESPA, we observed accuracy comparable to that of the backbone model, achieving 66.4\% of top-1 and 87.2\% of top-5 accuracy for the encrypted classification.
In contrast, the use of polynomial approximations results in 2.3--2.4\% drop in accuracy.

To evaluate the impact of our approach on precision, we tested a single conv2d operation with varying channel counts, up to 1024, combined with StoC, and analyzed the errors from FHE execution. The largest errors occurred when 1024 channels were packed per ciphertext, with both NeuJeans and \cite{kim_2023_optcnn} yielding mean absolute errors around 1.4e-6. Given that the error from ReLU polynomial approximation \cite{lee_2021_precise} is around 1.5e-4, the conv2d errors are minimal and do not significantly impact accuracy.
Therefore, the disparity in accuracy drop can be attributed to whether the original network operates as-is (using AESPA) or uses approximation. 
Prior FHE-based CNN inference studies~\cite{lee_2022_low,kim_2023_optcnn} have also experienced non-negligible accuracy drops due to errors in approximation. However, the impact appears more pronounced in our evaluation, likely due to the larger size of our network.

\subsection{Client Overhead}

\setlength{\tabcolsep}{4pt}
\begin{table}[t]
\centering
\caption{The cost of client-side tasks on a laptop computer for ImageNet inference. Client-side tasks stay the same regardless of the parameter set or the activation method.
}
\label{tab:client-cost}
\begin{tabular}{l|cccc}
\toprule
Implementation& \multicolumn{2}{c}{Encryption} & \multicolumn{2}{c}{Decryption}\\
(encoding, domain)& Time & Ctxt size & Time & Ctxt size\\
\midrule
\cite{kim_2023_optcnn} (coefficient, $\mathbb{R}^{N}$) & 17.3ms & {\color{white} 0}6MB & 2.11ms & 1MB\\
\name (CinS, $\mathbb{C}^{N/2}$)& 45.7ms & 12MB & 2.11ms & 1MB\\
\name (CinS, $\mathbb{R}^{N}$)& 22.1ms & {\color{white} 0}6MB& 2.11ms & 1MB\\
\bottomrule
\end{tabular}
\end{table}
\setlength{\tabcolsep}{6pt}

Although the client system used for the evaluation has much lower computational capability than the server system, client computation time accounts for only a tiny portion in the end-to-end inference time.
Table~\ref{tab:client-cost} shows the execution time of client-side tasks, which only include encryption and decryption.
Compared to \cite{kim_2023_optcnn} using Coefficient encoding, which has negligible encoding cost, \name involves additional computations for CinS encoding, adding 4.8ms to the encryption time.
It is important to use the dense packing for real numbers in \S\ref{sec:dense-pack} because it reduces the number of ciphertexts to encrypt for a fixed input size, also reducing the total encryption time.
For decryption, Slot encoding is used for all the implementations, resulting in the same decryption time.
Regardless of the implementation method, the use of FHE incurs small overhead for the clients, adding only dozens of milliseconds to the end-to-end inference time.

Table~\ref{tab:client-cost} also shows the total size of ciphertexts that are sent to and received from the server for a single inference.
The use of FHE enables \name to perform CNN inference with only 7MB of total data communication,

\section{Conclusion}

\name is an optimized solution for private inference of deep convolutional neural networks (CNNs) based on a cryptographic primitive, fully homomorphic encryption (FHE).
\name incorporates efficient algorithms for the evaluation of convolutional layers (conv2d), enabled by a dedicated encoding method for FHE ciphertexts.
Through leveraging common patterns in conv2d and bootstrapping, \name eliminates the computational redundancy by fusing conv2d with bootstrapping.
Finally, \name provides FHE-friendly execution flows that minimize the cost of bootstrapping and various conv2d layers when applying our methods to the private end-to-end inference of deep CNNs.
Our experiments exhibit up to 5.68$\times$ of speedups achieved by \name across various types of conv2d.
We demonstrate that \name performs ResNet18 ImageNet classification within 5.35 seconds on an A100 GPU system, while inducing minimal computation and communication overhead for the client.

\begin{acks}
This work was supported by Samsung Electronics Co., Ltd(IO201207-07812-01) and Institute of Information \& communications Technology Planning \& Evaluation (IITP) grant funded by the Korea government (MSIT) (No. 2021-0-01343, IITP-2023-RS-2023-00256081, and IITP-2024-RS-2024-00469698).
The ICT at Seoul National University (SNU) provides research facilities for this study.
Jung Ho Ahn, the corresponding author, is with the Department of Intelligence and Information and the Interdisciplinary Program in Artificial Intelligence, SNU.
\end{acks}









\bibliographystyle{ACM-Reference-Format}
\bibliography{reference}
\appendix

\section{Real C\lowercase{in}S Encoding}
\label{app:real_cins_encoding_pf}
Consider a real message vector $\mathbf{m} = (\mathbf{m}_1 | \mathbf{m}_2 | \cdots | \mathbf{m}_C)\in\mathbb{R}^{N}$ that is CinS encoded as in \S\ref{sec:dense-pack}. Then from the matrix properties of \S\ref{subsection:partialDFT}, the following holds.

\small
\begin{equation}
    \begin{split}
        \langle \mathbf{m} \rangle_\text{CinS} =& \langle S_{\ell \leftarrow 1}\mathcal{P}_\ell(\check{\mathbf{m}}_1 | \cdots | \check{\mathbf{m}}_C) \rangle_\text{slot}\\
        =& \langle S_{\ell \leftarrow 1}(P_{\ell}\check{\mathbf{m}}_1 | \cdots | P_{\ell}\check{\mathbf{m}}_C) \rangle_\text{slot}\\
        =& \langle (\mathcal{T}_{\ell}P_{\ell}\check{\mathbf{m}}_1 | \cdots | \mathcal{T}_{\ell}P_{\ell}\check{\mathbf{m}}_C) \rangle_\text{slot}
    \end{split}
\end{equation}
\normalsize

Let $DFT'_{2\ell}(\mathbf{m}_i)=\mathcal{T}_{\ell}P_{\ell}\check{\mathbf{m}}_i$ as in shown Equation~\ref{eq:nested_real_vector}. Since $P_{\ell}$ is an $\ell \times \ell$ bit-reversal permutation matrix, we can simplify $\mathcal{T}_\ell P_{\ell}$.

\small
\begin{equation}\label{eq:twiddle_factor}
\mathcal{T}_\ell P_{\ell} = \left( \omega_{4\ell}^{5^{j} \cdot {\sf rev}_{\ell}(k)}\right)_{0 \le j,k < \ell} P_{\ell}
= \left( \omega_{4\ell}^{5^{j} \cdot k}\right)_{0 \le j,k < \ell}
\end{equation}
\normalsize

$\omega_{4\ell}^{5^{j}}$ has the following property. 

\small
\begin{equation*}
(\omega_{4\ell}^{5^{j}})^{\ell}=e^{2\pi \sqrt{-1} \ell5^{j}/ 4\ell}=(e^{\pi \sqrt{-1}/2})^{5^{j}}={(\sqrt{-1})}^{5^{j}}=\sqrt{-1}
\end{equation*}
\normalsize

Then, $(DFT'_{2\ell}(\mathbf{m}_i))_j$ becomes

\small
\begin{equation*}
\begin{split}
&(DFT'_{2\ell}(\mathbf{m}_i))_j \\
=& (\left( \omega_{4\ell}^{5^{j} \cdot k}\right)_{0 \le j,k < \ell}\check{\mathbf{m}}_i)_j \\
=&\sum_{k=0}^{\ell-1}(\mathbf{m}_i)_k(\omega_{4\ell}^{5^{j}})^{k}+\sum_{k=0}^{\ell-1}\sqrt{-1}(\mathbf{m}_i)_{\ell+k}(\omega_{4\ell}^{5^{j}})^{k} \\
=&\sum_{k=0}^{\ell-1}(\mathbf{m}_i)_k(\omega_{4\ell}^{5^{j}})^{k}+\sum_{k=0}^{\ell-1}(\mathbf{m}_i)_{\ell+k}(\omega_{4\ell}^{5^{j}})^{\ell+k} \\
=&\sum_{k=0}^{2\ell-1}(\mathbf{m}_i)_k(\omega_{4\ell}^{5^{j}})^{k}
\end{split}
\end{equation*}
\normalsize

Given this definition of $DFT'_{2\ell}$, the following holds for real vectors $\mathbf{x}, \mathbf{y} \in \mathbb{R}^{2\ell}$ ($*_{2\ell}$ is the negacyclic convolution).

\small
\begin{equation*}
    \begin{split}
        &(DFT'_{2\ell}(\mathbf{x}) \odot DFT'_{2\ell}(\mathbf{y}))_j\\
        =& (DFT'_{2\ell}(\mathbf{x}))_j \cdot (DFT'_{2\ell}(\mathbf{y}))_j\\
        =& (\sum_{k=0}^{2\ell-1}(\mathbf{x})_{k}(\omega_{4\ell}^{5^{j}})^{k}) \cdot (\sum_{k=0}^{2\ell-1}(\mathbf{y})_{k}(\omega_{4\ell}^{5^{j}})^{k})\\
        =& \sum_{k=0}^{2\ell-1}\sum_{s=0}^{k}(\mathbf{x})_{s}(\mathbf{y})_{k-s}(\omega_{4\ell}^{5^{j}})^{k}
        + \sum_{k=2\ell}^{4\ell-1}\sum_{s=k-2\ell+1}^{2\ell-1}(\mathbf{x})_{s}(\mathbf{y})_{k-s}(\omega_{4\ell}^{5^{j}})^{k}\\
        =& \sum_{k=0}^{2\ell-1}\sum_{s=0}^{k}(\mathbf{x})_{s}(\mathbf{y})_{k-s}(\omega_{4\ell}^{5^{j}})^{k}
        + \sum_{k=0}^{2\ell-1}\sum_{s=k+1}^{2\ell-1}(\mathbf{x})_{s}(\mathbf{y})_{2\ell+k-s}(\omega_{4\ell}^{5^{j}})^{k}(\omega_{4\ell}^{5^{j}})^{2\ell}\\
        =& \sum_{k=0}^{2\ell-1}\left(\sum_{s=0}^{2\ell-1}(-1)^{\lfloor\frac{k-s}{2\ell}\rfloor}(\mathbf{x})_{s}(\mathbf{y})_{(k-s)\mathsf{mod}{2\ell}}\right)(\omega_{4\ell}^{5^{j}})^{k}\\
        =& \sum_{k=0}^{2\ell-1}(\mathbf{x} *_{2\ell} \mathbf{y})_k(\omega_{4\ell}^{5^{j}})^{k}\\
        =& (DFT'_{2\ell}(\mathbf{x} *_{2\ell} \mathbf{y}))_j
    \end{split}
\end{equation*}
\normalsize

Thus we can derive the core property of real CinS encoding.

\small
\begin{equation*}
\begin{split}
&\langle \mathbf{m} \rangle_\text{CinS} \cdot \langle \mathbf{m}' \rangle_\text{CinS}\\
= &\langle(DFT'_{2\ell}(\mathbf{m}_1) | \cdots | DFT'_{2\ell}(\mathbf{m}_C))
\odot (DFT'_{2\ell}(\mathbf{m}'_1) | \cdots | DFT'_{2\ell}(\mathbf{m}'_C))\rangle_\text{slot}\\
= &\langle(DFT'_{2\ell}(\mathbf{m}_1*_{2\ell}\mathbf{m}'_1) | \cdots | DFT'_{2\ell}(\mathbf{m}_C*_{2\ell}\mathbf{m}'_C))\rangle_\text{slot}\\
= &\langle ({\mathbf{m}_1 *_{2\ell} \mathbf{m}'_1} | \cdots | {\mathbf{m}_{C} *_{2\ell} \mathbf{m}'_{C}}) \rangle_\text{CinS}
\end{split}
\end{equation*}
\normalsize

\begin{figure}
\centering
\includegraphics[width=.9\columnwidth]{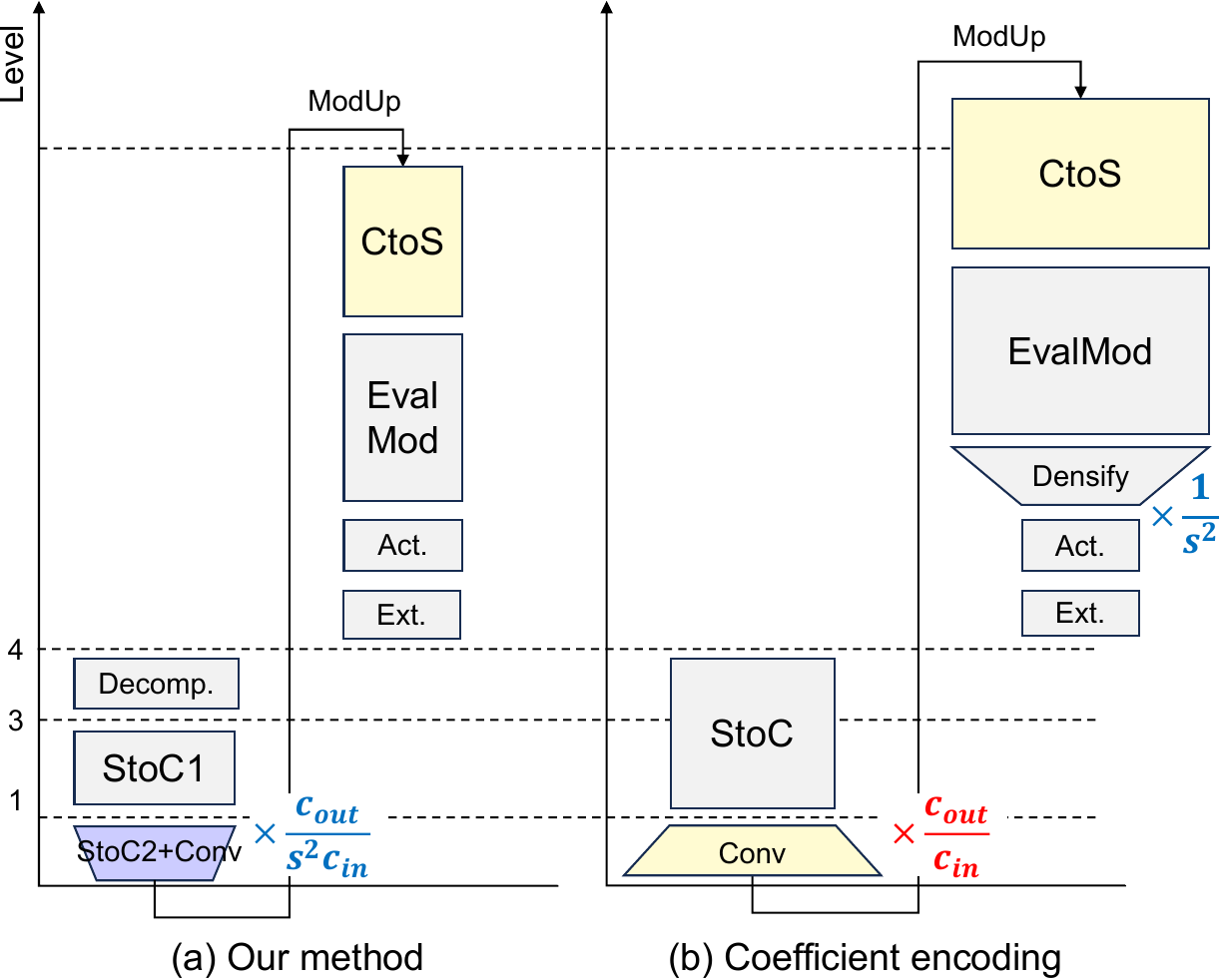}
\caption{Levels of ciphertexts when evaluating downsampling conv2d and activation along with boostrapping. The width of a block indicates the number of ciphertexts in each operation of a ResNet downsampling layer.}
\label{fig:jobScheduling}
\end{figure}
\section{Downsampling Conv2d}
\label{appendix:permutation_matrix}
We explain the decompose/densify methods from \S\ref{sec:flow:downsample} in more detail. Both decompose and densify operations require the image to be packed sparsely in slots, and execute fine-grained masking (PMult) and rotation (HRot). However, decompose prevents the mass generation of unwanted result by changing the image and kernel arrangement before evaluating conv2d, while densify removes unwanted result after it has been created. The computation of bootstrapping after single-stride conv2d, and before densify, greatly amplifies this difference, as shown in Figure~\ref{fig:jobScheduling}(b).

Computationally, decompose and densify are surprisingly similar. This is because the pattern of the data which densify extracts is almost identical to the rearrangement pattern of decompose. In detail, the densify operation extracts and rearranges only the red colored data from Figure~\ref{fig:sconv-decomposition}. Decompose is essentially repeating the densify operation (HMatmul$(\mathcal{G}_\ell, \cdot)$) four times, but with rotations before and after (line 5, 6 of Algorithm~\ref{alg:conv2d_downsample}) for applying densify to the right spot, and combining the results.

The exact form of matrix $\mathcal{G}_\ell$ is rather complex because the arrangement of the pixels in Figure~\ref{fig:sconv-decomposition} undergoes bit-reversal permutation when being converted between different encodings. Here we write a brief description of the matrix $\mathcal{G}_\ell$ from Algorithm~\ref{alg:conv2d_downsample}, and refer the reader to ~\cite{kim_2023_optcnn} for a more detailed explanation.

$\mathcal{G}_\ell$ is a block diagonal matrix of total size $N/2 \times N/2$. Each block is a $\ell \times \ell$ matrix $G_\ell$. The matrix $G_\ell$ itself is also a $2W \times 2W$ block matrix $((G_\ell)_{i, j})_{0\le i, j<2W}$, with each block being a matrix of size $W/4 \times W/4$. $(G_\ell)_{i, j}$ can be expressed as the following, where $I_n$ is an $n \times n$ identity matrix and $O_n$ is an $n \times n$ zero matrix.

\begin{equation*}
    (G_\ell)_{i, j} = \begin{cases}
        I_{W/4} & \text{if $j = 2i$ and $i \le W/2$} \\
        O_{W/4} & \text{otherwise}
    \end{cases}
\end{equation*}

\begin{table}[tb!]

\renewcommand{\arraystretch}{1.07} 
    \centering
    \caption{Notations Summary}
    \begin{tabular}{l p{.76\columnwidth}}
    \toprule
        Notation & Description \\
    \midrule
        $N$ & Degree of plaintext polynomial. \\
        $\ell$ & Length of a vector slice, which can be any power-of-two smaller than $N$. \\
        $C$ & Number of vector slices that fit in a message. $N/\ell$ for real CinS encoding and $(N/2)/\ell$ for complex. \\
        $P_{\ell}$ & Length $\ell$ bit-reversal permutation matrix. \\
        $\mathcal{P}_\ell$ & Block diagonal matrix of a total size of $N/2 \times N/2$, where each block is $P_\ell$. \\
        $\mathcal{P}$ & $\mathcal{P}_{N/2}$. Length $N/2$ bit-reversal permutation matrix. \\
	$\mathcal{T}_{n}$ & Length $n$ DFT matrix with bit-reversal permutation. \\
        $S_k$ & $\mathcal{T}_{N/2} = S_{N/4} \cdot S_{N/8} \cdots S_1$.\\
	$S_{j \leftarrow i}$ & $S_{j/2}S_{j/4}\cdots S_{i}$. $S_{N/2 \leftarrow 1} = \mathcal{T}_{N/2}$. \\
        $\mathbf{\bhat{x}}$ & $DFT_{\ell} (\mathbf{x})$, where $\mathbf{x}\in\mathbb{C}^{\ell}$. \\
        $\check{\mathbf{x}}$ & $(\check{\mathbf{x}})_i := (\mathbf{x})_i+\sqrt{-1}(\mathbf{x})_{i+\ell}$, where $\mathbf{x}\in\mathbb{R}^{2\ell}$. \\
        $\odot$ & Element-wise multiplication. \\
        $*_{\ell}$ & Length $\ell$ (negacyclic) convolution. \\
        $*_\text{local}$ & $\mathbf{m} *_\text{local} \mathbf{m}' := 
(\mathbf{m}_{0} *_{\ell} \mathbf{m}'_{0} | \cdots | \mathbf{m}_{C-1} *_\ell \mathbf{m}_{C-1}^\prime)$. \\
        $\langle \mathbf{m} \rangle_{type}$ & Plaintext encoding $\mathbf{m}$, using $type$ encoding. \\
        $[\mathbf{pt}]$ & Ciphertext encrypting plaintext $\mathbf{pt}$.\\
        $s$ & Stride of a strided conv2d. \\
        $C_{in}$ & Number of input channels in conv2d. \\
        $C_{out}$ & Number of output channels in conv2d. \\
    \bottomrule
    \end{tabular}
    \label{tab:symbol_table}
\end{table}

\end{document}